\documentclass[fleqn,usenatbib]{mnras}

\usepackage{newtxtext,newtxmath}

\usepackage[T1]{fontenc}
\usepackage{ae,aecompl}

\usepackage{graphicx}	
\usepackage{amsmath}	
\usepackage{amssymb}	
\usepackage{longtable}
\usepackage{multirow}

\title[Morpho-kinematic properties of S0 bulges]{Morpho-kinematic properties of field S0 bulges in the CALIFA survey}

\author[J. M\'endez-Abreu]{J. M\'endez-Abreu$^{1,2,3}$\thanks{E-mail: jairomendezabreu@gmail.com},
J. A. L. Aguerri$^{2,3}$,
J. Falc\'on-Barroso$^{2,3}$,
T. Ruiz-Lara$^{2,3,4,5}$,
\newauthor L. S\'anchez-Menguiano$^{6,4}$,
A. de Lorenzo-C\'aceres$^{1,4,15}$,
L. Costantin$^{7}$,
\newauthor C. Catal\'an-Torrecilla$^{8}$,
L. Zhu$^{9}$,
P. S\'anchez-Blazquez$^{10,11}$,
E. Florido$^{4,5}$,
E. M. Corsini$^{7,12}$,
\newauthor V. Wild$^{1}$,
M. Lyubenova$^{13}$,
G. van de Ven$^{9}$,
S. F. S\'anchez$^{15}$,
J. Bland-Hawthorn$^{16}$,
\newauthor L. Galbany$^{17,18}$,
R. Garc\'ia-Benito$^{6}$,
B. Garc\'ia-Lorenzo$^{2,3}$,
R. M. Gonz\'alez Delgado$^{6}$,
\newauthor A. R. L\'opez-S\'anchez$^{19}$,
R. A. Marino$^{20}$,
I. M\'arquez$^{6}$,
B. Ziegler$^{21}$,
\newauthor and the CALIFA collaboration
\\
$^{1}$School of Physics and Astronomy, University of St. Andrews, SUPA, North Haugh, KY16 9SS, St. Andrews, UK\\
$^{2}$Instituto de Astrof\'isica de Canarias, Calle V\'ia L\'actea s/n, E-38205 La Laguna, Tenerife, Spain\\
$^{3}$Departamento de Astrof\'isica, Universidad de La Laguna, E-38200 La Laguna, Tenerife, Spain\\
$^{4}$Departamento de F\'isica Te\'orica y del Cosmos, Universidad de Granada, Campus de Fuentenueva, E-18071 Granada, Spain\\
$^{5}$Instituto Carlos I de F\'isica Te\'orica y Computacional, Universidad de Granada, E-18071 Granada, Spain\\
$^{6}$Instituto de Astrof\'isica de Andaluc\'ia (CSIC), Glorieta de la Astronom\'ia s/n, E-3004, E-18080 Granada, Spain\\
$^{7}$Dipartimento di Fisica e Astronomia 'G. Galilei', Universit\'a di Padova, vicolo dell'Osservatorio 3, I-35122 Padova, Italy\\
$^{8}$Departamento de Astrof\'isica y CC. de la Atm\'osfera, Universidad Complutense de Madrid, E-28040 Madrid, Spain\\
$^{9}$Max Planck Institute for Astronomy, K\"onigstuhl 17, D-69117 Heidelberg, Germany\\
$^{10}$Departamento de F\'isica Te\'orica, Universidad Aut\'onoma de Madrid, E-28049 Cantoblanco, Spain\\
$^{11}$Instituto de Astronom\'ia, Pontificia Universidad Cat\'olica de Chile, Avenida Vicu\~na Mackenna 4860, Macul, Santiago, Chile\\
$^{12}$INAF-Osservatorio Astronomico di Padova, vicolo dell'Osservatorio 5, I-35122 Padova, Italy\\
$^{13}$Kapteyn Astronomical Institute, University of Groningen, PO Box 800, NL-9700 AV Groningen, the Netherlands\\
$^{14}$Max Planck Institute for Astronomy, K\"onigstuhl 17, D-69117 Heidelberg, Germany\\
$^{15}$Instituto de Astronom\'ia, Universidad Nacional Aut\'onoma de M\'exico, A.P. 70-264, 04510 M\'exico D.F., Mexico\\
$^{16}$Sydney Institute for Astronomy, School of Physics A28, University of Sydney, NSW 2006, Australia\\
$^{17}$Millennium Institute of Astrophysics, Chile\\
$^{18}$Departamento de Astronom\'ia, Universidad de Chile, Camino El Observatorio 1515, Las Condes, Santiago, Chile\\
$^{19}$Australian Astronomical Observatory, PO BOX 296, Epping, 1710 NSW, Australia\\
$^{20}$Department of Physics, Institute for Astronomy, ETH Z\"urich, CH-8093 Z\"urich, Switzerland\\
$^{21}$University of Vienna, T\"urkenschanzstrasse 17, 1180 Vienna, Austria\\
}

\date{Accepted XXX. Received YYY; in original form ZZZ}

\pubyear{2017}

\begin{document}
\label{firstpage}
\pagerange{\pageref{firstpage}--\pageref{lastpage}}
\maketitle

\begin{abstract}

We study a sample of 28  S0 galaxies extracted from the integral-field
spectroscopic   (IFS)  survey   CALIFA.    We   combine  an   accurate
two-dimensional  (2D) multi-component  photometric decomposition  with
the IFS  kinematic properties  of their  bulges to understand their
formation scenario.   Our final sample  is representative of  S0s with
high  stellar  masses ($M_{\star}  /M_{\sun}  >  10^{10}$).  They  lay
mainly  on   the  red  sequence   and  live  in   relatively  isolated
environments similar to that of the field and loose groups.

We  use  our 2D  photometric  decomposition  to  define the  size  and
photometric properties of the bulges, as well as their location within
the galaxies.  We perform mock spectroscopic simulations mimicking our
observed galaxies to quantify the impact of the underlying disc on our
bulge kinematic  measurements ($\lambda$ and $v/\sigma$).   We compare
our  bulge  corrected kinematic  measurements  with  the results  from
Schwarzschild dynamical  modelling.  The  good agreement  confirms the
robustness  of our  results and  allows  us to  use bulge  deprojected
values of $\lambda$ and $v/\sigma$.  We find that the photometric ($n$
and $B/T$) and kinematic ($v/\sigma$  and $\lambda$) properties of our
field  S0  bulges  are  not  correlated.   We  demonstrate  that  this
morpho-kinematic decoupling is  intrinsic to the bulges and  it is not
due to projection effects.

We conclude  that photometric diagnostics to  separate different types
of  bulges  (disc-like  vs  classical)  might not  be  useful  for  S0
galaxies.  The  morpho-kinematics properties  of S0 bulges  derived in
this  paper  suggest that  they  are  mainly formed  by  dissipational
processes happening  at high  redshift, but dedicated  high-resolution
simulations are necessary to better identify their origin.

\end{abstract}



\section{Introduction}
\label{sec:intro}

The  Hubble  tuning fork  diagram  \citep{hubble36}  has provided  for
decades the benchmark to study  galaxy evolution. In recent years, the
Hubble  diagram has  been  revisited a  number of  times  in order  to
accommodate new  photometric and kinematic properties  of the galaxies
\citep{cappellari11,kormendybender12}.     Most   of    the   proposed
modifications affect the position of  lenticular galaxies (S0s) in the
diagram.   S0  galaxies  were  initially placed  at  the  intersection
between  ellipticals   and  spirals,  implying  that   they  formed  a
homogeneous   class   of  galaxies.    Since   the   early  works   by
\citet{vandenbergh76} this  homogeneity has  been discarded,  but only
now it is  commonly accepted that they encompass a  complete family of
galaxies  representing  a  distinct  branch  of  the  Hubble  diagram.
Therefore, understanding the origin of lenticular galaxies and whether
they are related to spiral or elliptical galaxies is still a challenge
for contemporary astrophysics \citep[see][for a review]{aguerri12}.

The bulge prominence, or relative size with respect to the galaxy, has
been  one  of  the  primary  features used  to  classify  galaxies  in
different  Hubble  types.   However,  defining what  a  bulge  is  not
straightforward.   Historically,  a  bulge  was defined  as  a  bright
central  concentration  due  to  stellar  light  with  relatively  few
features due to  dust and star formation  \citep{hubble36}.  This {\it
  morphological} definition was extensively  used to produce a variety
of   visual  classification   schemes  for   galaxies  \citep[see][and
  references  therein]{buta13}.    With  the  advent   of  photometric
decompositions, a more quantitative  definition naturally arose.  This
{\it photometric} definition considers the bulge as the extra light in
the central region of the disc,  above the inwards extrapolation of an
exponential   disc  \citep{freeman70}.    Nowadays,  the   photometric
definition of a  bulge is widely used, and it  has been generalised to
the  central  bright  structure,  usually described  with  a  S\'ersic
profile \citep{sersic68}, prevailing amongst  other structures such as
discs,   bars,    lenses,   etc,   in    multi-component   photometric
decompositions          \citep{gadotti09,laurikainen10,mendezabreu14}.
Throughout this paper we use the  photometric definition of a bulge in
order to compare with the literature.

The  structure of  S0  galaxies  is an  example  of their  complexity.
Despite  initially  being classified  as  systems  with only  a  bulge
dominating the  light at the galaxy  centre and an outer  disc without
indication  of spiral  arms, recent  works have  provided a  wealth of
evidence   for  multiple   structures:   bars,   lenses,  rings,   etc
\citep[e.g.,][]{laurikainen13}.  Still, the  bulge prominence, usually
characterised by its luminosity ratio with respect to the whole galaxy
light  ($B/T$), is  considered the  main parameter  to morphologically
classify  different   S0  galaxies  \citep[i.e.,][]{kormendybender12}.
There is ample observational evidence that bulges in S0 galaxies cover
a  wide range  of physical  properties such  as $B/T$,  S\'ersic index
($n$),  rotational  support,  and   even  stellar  populations.   This
supports a scenario  were different types of bulges can  be present at
the            centre             of            S0            galaxies
\citep{delorenzocaceres12,mendezabreu14,erwin15}.

The  current observational  picture of  galactic bulges  divides these
systems  into  two  broad  classes:  classical  and  disc-like  bulges
\citep{kormendykennicutt04,athanassoula05}.
An updated list  of the observational criteria to  separate both types
of  bulges is  given  in \citet{fisherdrory16}.   In short,  disc-like
bulges  are  usually   oblate  ellipsoids  \citep{mendezabreu10b}  with
apparent   flattening    similar   to   their   outer    discs,   with
surface-brightness distributions  well fitted with a  S\'ersic profile
of index $n<2$ \citep{fisherdrory08}, and $B/T<0.35$. Their kinematics
are  dominated by  rotation  in  diagrams such  as  the $v/\sigma$  vs
$\epsilon$   \citep{kormendykennicutt04}  and   thus  they   are  also
low-$\sigma$     outliers     of    the     Faber-Jackson     relation
\citep{faberjackson76}.  Disc-like  bulges are also  usually dominated
by young  stars, with  the presence  of gas  and possible  recent star
formation \citep{fisherdrory16}.  On the  other hand, classical bulges
are thought to follow surface-brightness distributions with a S\'ersic
index $n>2$ and  $B/T>0.5$, they appear rounder  than their associated
discs, and  their stellar kinematics  are dominated by  random motions
that  generally   satisfy  the  fundamental  plane   (FP)  correlation
\citep{bender92, falconbarroso02,aguerri05}.   The stellar populations
of classical bulges show similarities with those of ellipticals of the
same  mass. In  general,  they are  old and  metal-rich  with a  short
formation  timescale   \citep[see][for  a  review  on   their  stellar
  populations]{sanchezblazquez16}.   Nevertheless,  this dichotomy  of
the observed  properties is  still controversial since  recent studies
claim the  different properties of  bulges can  be just driven  by the
bulge mass \citep{costantin17}.

Different  formation  scenarios  have  been proposed  to  explain  the
observational differences between classical  and disc-like bulges. The
former can  be created via  dissipative collapse of  protogalactic gas
clouds  \citep{eggen62}  or by  the  coalescence  of giant  clumps  in
primordial discs  \citep{noguchi99,bournaud07}.  Moreover,  they could
also  grow out  of  disc material  externally  triggered by  satellite
accretion during minor  merging events \citep{aguerri01,elichemoral06}
or  by  galaxy  mergers   \citep{kauffmann96}  with  different  merger
histories \citep{hopkins09}.   Disc-like bulges are thought  to be the
products     of      secular     processes     driven      by     bars
\citep[][]{kormendykennicutt04}.  Bars are ubiquitous in disc galaxies
\citep[e.g.,][]{eskridge00,aguerri09}.  They  are efficient mechanisms
for driving gas inward to  the galactic centre triggering central star
formation generally  associated with disc-like  bulges.  Nevertheless,
\citet{elichemoral11}  have recently  proposed  that disc-like  bulges
might  also  be  created  by  the  secular  accretion  of  low-density
satellites into the main galaxy,  thus providing an alternative to the
bar-driven growth  of disc-like  bulges.  Understanding the  nature of
bulges of S0s  in the nearby Universe would  set important constraints
on models of S0 formation and evolution.

The non-homogeneity  of the S0  family of  galaxies has also  raised a
number  of  new  formation  theories   to  explain  their  variety  of
properties. One of the most  commonly proposed formation scenarios for
S0 galaxies  suggests that they  are descendants from  spiral galaxies
that happen to quench  their star formation \citep{bekkicouch11}.  The
mechanism responsible  for this  transformation has  to stop  the star
formation in the  disc and enhance the  spheroidal component.  Several
physical processes  have been  invoked to  produce these  two effects,
most  of them  directly related  to the  presence of  the galaxy  in a
high-density environment.   To enhance  the spheroidal  component, the
harassment scenario proposes that the cumulative effects of fast tidal
encounters  between  galaxies  and   with  the  cluster  gravitational
potential  can  produce   dramatic  morphological  transformations  in
galaxies   \citep{bekki98,   moore98,  moore99,governato09}.    Galaxy
harassment  in clusters  \citep{moore96}  is able  to  remove a  large
amount of  mass from both  the disc and halo,  but not from  the bulge
where     the      stars     are     more      gravitational     bound
\citep{aguerrigonzalezgarcia09}.  Stopping  the star formation  of the
disc involves either the direct stripping of cold gas from the disc of
the galaxy \citep[e.g.,  ram pressure,][]{gunngott72,quilis00}, or the
removal  of  its  hot  halo  gas  reservoir  over  a  long  period  of
strangulation  \citep{larson80,   balogh00}.   These   mechanisms  act
preferentially on gas, causing little or no disruption to the galaxy's
stellar disc, but they need different time scales.

Interestingly,  S0  galaxies  are  found  in  all  environments,  from
high-density  clusters  to  the  field,  allowing  for  a  variety  of
evolutionary paths that are not related with high-density environments
\citep{wilman09,bekkicouch11}.   Galaxy mergers  are one  of the  most
widely  studied mechanisms  which  show the  potential  to form  S0s.
Recently,  \citet{querejeta15b} used  the Calar  Alto Legacy  Integral
Field Area survey \citep[CALIFA,][]{sanchez12}  data to prove that the
stellar  angular  momentum  and   concentration  of  late-type  spiral
galaxies are  incompatible with those  of S0s, therefore  suggesting a
merger origin for S0 galaxies.  However, stellar discs of galaxies are
typically   disrupted   by   these   processes,   requiring   specific
environmental conditions for disc survival \citep{hopkins09} or a long
period of disc regrowth  from the surrounding gas \citep{kannappan09}.
In the merger paradigm, the central bulge of disc galaxies forms prior
to the  disc as  a result  of early  merging. Despite  this inside-out
formation   scenario   is    compatible   with   recent   observations
\citep{gonzalezdelgado15},  the   amount  of  gas  available   in  the
progenitor  galaxies  has been  shown  to  be  a  clue for  the  bulge
evolution,  with dissipative  processes driving  the consequent  bulge
growth  rather  than  the   redistribution  of  stars  \citep[see][and
  references  therein]{brooks16}.  At  lower  redshift, minor  mergers
might have  a higher  incidence in galaxy  evolution than  major ones.
The final remnant disc is usually compatible with that of S0 galaxies,
and multiple  merging with small  satellites can produce  bulge growth
mimicking  the  properties   of  S0s  \citep{aguerri01,elichemoral06}.
Galaxy fading can also occur  through internal secular processes.  One
of the  internal processes that  could regulate the star  formation in
galaxies is feedback  processes, due to supernovae  or active galactic
nuclei (AGN), heating the cold gas in galaxies and stopping their star
formation.  This  process would  transform early-type  spiral galaxies
located in the blue cloud into galaxies located close to or in the red
sequence   \citep[][]{schawinski06}.   These   transformed  early-type
galaxies could be the progenitors of later S0 galaxies.

In  this  paper,  we  have   studied  the  photometric  and  kinematic
properties of a  well defined sample of 28 S0  galaxies extracted from
the  CALIFA   survey  \citep{sanchez12}.   The   accurate  photometric
decomposition carried out by \citet{mendezabreu17} using the $g-, r-,$
and  $i-$  bands provided  by  the  Sloan  Digital Sky  Survey  (SDSS)
combined  with  the high  quality  integral  field spectroscopy  (IFS)
obtained from CALIFA  have allowed us to characterise  these bulges to
an  unprecedented   level  of  detail.    Our  main  emphasis   is  to
characterise  the morpho-kinematics  properties of  S0 bulges  to shed
light on their possible formation scenarios.

The  careful  selection of  a  sample  of {\it  bona-fide}  lenticular
galaxies is key in this work.  Therefore, we have developed a detailed
methodology that allows us to deal with the well-known difficulties of
separating early-type galaxies into  ellipticals and lenticulars using
only photometric data.  Our final aim  is to find a sample of galaxies
that  can   be  photometrically  well   described  by,  at   least,  a
two-component  model (bulge  and  disc) in  the  {\it canonical  way}.
These  galaxies  have  therefore   an  inner  photometric  bulge  that
dominates   only  the   central  parts   of  the   surface  brightness
distribution and a disc dominating  the light in the galaxy outskirts.
Further  structural  components  such   as  bars  or  truncated  outer
profiles, not expected in elliptical  galaxies, are also signatures of
a photometric lenticular galaxy.  The  process described in this paper
implies that some lenticular galaxies will be erroneously removed from
the  analysis, but  we prefer  to work  with a  safe and  well-defined
sample of photometric S0 galaxies.

The paper  is organized  as follows:  Sect.~\ref{sec:sample} describes
the   initial   sample   of   early-type   galaxies   used   in   this
work. Sect.~\ref{sec:surphot}  details the analysis of  the early-type
galaxies    surface-brightness     distribution.     In    particular,
Sect.~\ref{sec:S0sample} presents the methodology followed to separate
elliptical and lenticular galaxies  from our initial early-type galaxy
sample.   This  analysis will  be  used  for  the final  selection  of
photometrically  defined  lenticular  galaxies  and  their  structural
analysis.  Sect.~\ref{sec:globalprop} describes the general properties
of our {\it  bona-fide} sample of lenticular  galaxies.  The kinematic
measurements using the CALIFA database,  as well as the correction due
to       disc       contamination,        are       described       in
Sect.~\ref{sec:kinmeasurements}.  Sect.~\ref{sec:results} presents the
main results  of this paper that  will be discussed in  the context of
bulge  formation in  Sect.~\ref{sec:discussion}.  The  conclusions are
given in Sect.~\ref{sec:conclusions}.  Throughout  the paper we assume
a flat cosmology with $\Omega_{\rm m}$ = 0.3, $\Omega_{\rm \Lambda}$ =
0.7, and a Hubble constant $H_0$ = 70 km s$^{-1}$ Mpc$^{-1}$.

\section{CALIFA sample of early-type galaxies}
\label{sec:sample}

This work  is based on  the observations taken  as part of  the CALIFA
survey   \citep{sanchez12}.     The   CALIFA   final    data   release
\citep[DR3,][]{sanchez16}  is   composed  by  two  different   set  of
galaxies: i) galaxies extracted from  the CALIFA mother sample and the
CALIFA extended sample.   The former was drawn from  the Sloan Digital
Sky   Survey  Data   Release   7   (SDSS-DR7)  photometric   catalogue
\citep{abazajian09}.   It is  composed  by 939  galaxies with  angular
isophotal diameter $45 < D_{25} <  80$ arcsec in $r-$band and within a
redshift range  $0.005 < z  < 0.03$.   The detailed properties  of the
mother  sample are  extensively discussed  in \citet{walcher14}.   The
extended sample  is a  compendium of galaxies  observed with  the same
CALIFA setup  as ancillary  science projects.  As  part of  the CALIFA
sample  characterisation, every  galaxy in  both samples  was visually
classified into  its corresponding  Hubble type independently  by five
members  of  the  collaboration.   The initial  sample  of  early-type
galaxies  used in  this  paper was  based  on the  mean  value of  the
morphological  type  derived  in this  classification  \citep[see][for
  details]{walcher14}.  In particular, we selected only those galaxies
with Hubble type ranging from ellipticals to S0 galaxies.

The  CALIFA observational  strategy  includes  observing every  galaxy
using two different setups.  The V500 grating has a nominal resolution
of $R$  = 850  at 5000\AA\,  and covers  from 3745\AA\,  to 7300\AA\,.
This grating  is particularly suitable for  stellar population studies
and  it has  been  extensively used  within  the CALIFA  collaboration
\citep[i.e.,][]{perez13,cidfernandes13,gonzalezdelgado14a,gonzalezdelgado14b,sanchezblazquez14,martinnavarro15,gonzalezdelgado15,gonzalezdelgado16,sanchez16}
and  for  studies  of  the  physical properties  of  the  ionized  gas
\citep[i.e.,][]{kehrig12,marino13,singh13,sanchez13,papaderos13,sanchez14,galbany14,wild14,sanchez15,garcialorenzo15,barreraballesteros15,holmes15,catalantorrecilla15,marino16,sanchezmenguiano16}.
The second  setup is based on  the V1200 grating with  better spectral
resolution  $R$  =  1650  at  4500\AA\,.   This  grating  covers  from
3400\AA\, to  4750\AA\, and is  perfectly suited to  kinematic studies
using stellar  absorption features  \citep[examples of its  use within
  the                       CALIFA                       collaboration
  includes][]{barreraballesteros14,aguerri15}.  In  this work,  we are
interested in the  kinematic properties of the bulges  in S0 galaxies,
therefore our initial sample is constrained to those galaxies observed
with the  V1200 grating.  After  removing those systems  undergoing an
interaction, with strongly disturbed morphologies, or highly inclined,
we  end up  with an  initial sample  of 81  early-type galaxies.   The
photometric properties  of these galaxies  were analysed in  detail in
\citet{mendezabreu17}.

\section{Surface photometry and S0 selection}
\label{sec:surphot}

The accurate analysis of  the surface-brightness distribution (SBD) of
our S0 galaxies is a critical step  in our study. First, it is used to
properly define  a sample of  {\it bona-fide} photometric  S0 galaxies
(Sect.~\ref{sec:S0sample}); second, the bulge size provides the galaxy
region   from    which   the   stellar   kinematics    are   extracted
(Sect.~\ref{sec:kinprop});  and  finally, the  structural  parameters,
combined  with  the  galaxy  kinematics, are  used  to  constrain  the
formation scenarios of S0 galaxies (Sect.~\ref{sec:discussion}).

The CALIFA DR3 sample is based  on the SDSS-DR7 database and therefore
high-quality, homogeneous, and multi-wavelength  imaging of the galaxy
sample is assured.  We used the  imaging frames in the $g$-, $r$-, and
$i$-bands provided in  the SDSS-DR7 to perform  our surface brightness
analysis \citep[see][]{mendezabreu17}.   These images  are pre-reduced
but they still contain information about the local sky background.  To
guarantee an accurate  analysis, we used our own  procedures to remove
the  sky background  instead  of  using the  tabulated  values in  the
SDSS-DR7 database (Sect.~\ref{sec:sdssimages}).

\subsection{SDSS images sky subtraction}
\label{sec:sdssimages}

\begin{figure}
\includegraphics[width=0.49\textwidth]{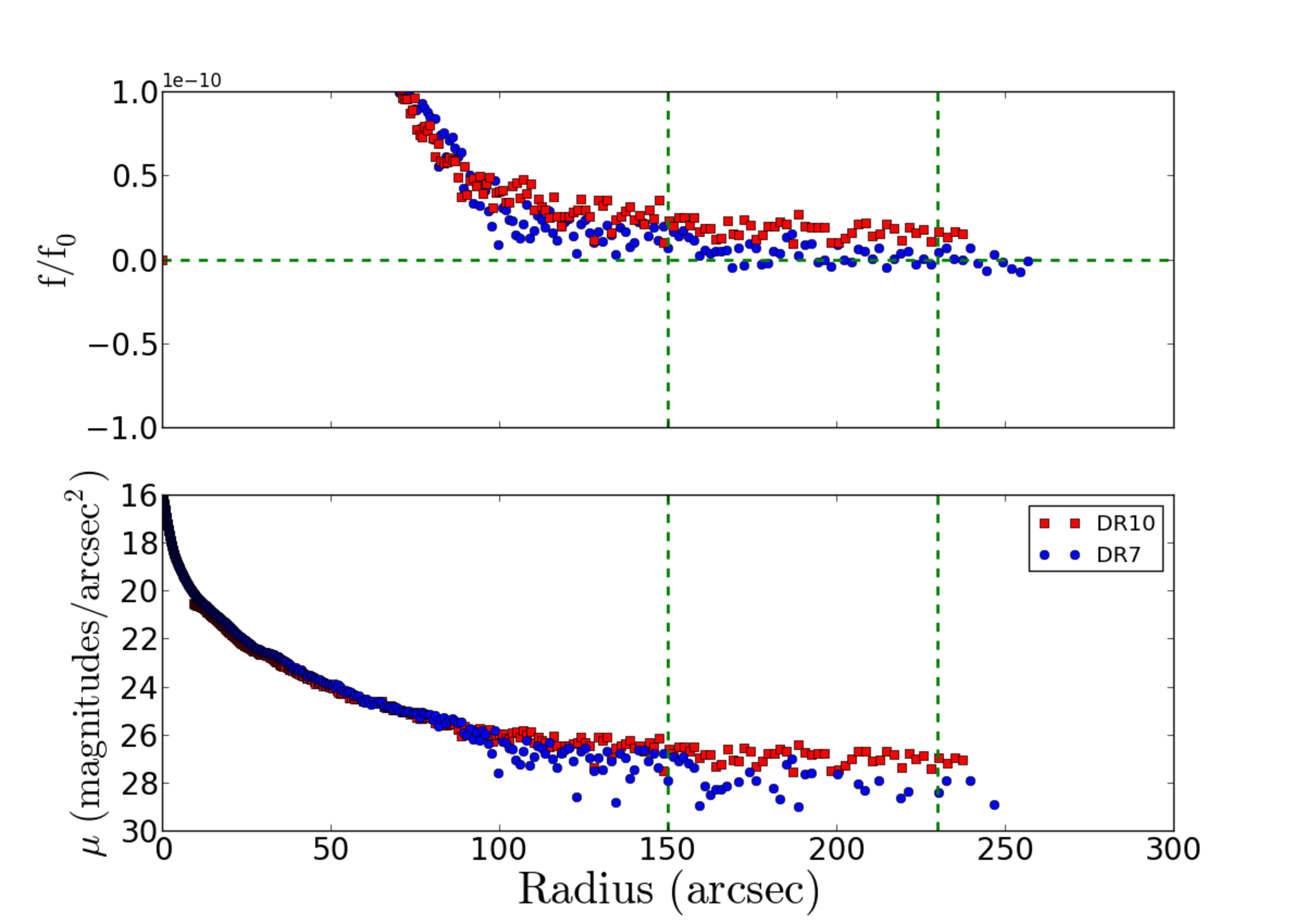}
\caption{Radial distribution of the  zero-point count rate ($f/f{_0}$)
  and  surface-brightness   profile  for   SDSS  $i-$band   using  our
  methodology on the DR7 and that  provided by the DR10 for the galaxy
  NGC0001. The vertical lines represent the region where the sky level
  was computed.   Note the improvement  of $\sim$  60 $\%$ in  the sky
  level comparing our sky subtraction scheme with the DR10 implemented
  one.}
\label{DR7_DR10}
\end{figure}

Although  SDSS-DR7 provides  a measurement  of the  sky level  (as the
median value of  every pixel after a sigma-clipping  is applied), this
estimate  has proven  insufficient,  specially to  study the  faintest
parts of spiral  galaxies \citep[][]{pohlentrujillo06}.  Therefore, we
apply  our  own  sky  subtraction  procedure  to  the  SDSS-DR7  fully
calibrated  frames.  We  automatically  mask out  foreground stars  in
every      frame     using      the     code      SOURCE     EXTRACTOR
\citep[SExtractor,][]{bertinarnouts96} as well as visually-inspect and
manually-mask small  features that SExtractor might  have missed. This
mask  will   also  be  provided   as  input  to  the   2D  photometric
decomposition     algorithm     at      a     later     stage     (see
Sect.~\ref{sec:photdec}).   We follow  the  sky subtraction  procedure
proposed  by  \citet[][]{pohlentrujillo06}:  first, we  use  the  {\it
  ellipse}  IRAF\footnote{{\tt IRAF}  is distributed  by the  National
  Optical Astronomy Observatory, which  is operated by the Association
  of Universities  for Research in Astronomy  (AURA) under cooperative
  agreement with the National Science  Foundation.} task to obtain the
one  dimensional   (1D)  surface-brightness  profile  using   a  fixed
ellipticity ($\epsilon$)  and position angle ($PA$)  matching the disc
outermost isophotes.   Figure~\ref{DR7_DR10} shows an example  of this
methodology  where $f_0$  is the  zero-point count  rate necessary  to
calibrate            the           SDSS            data\footnote{Check
  \url{http://www.sdss2.org/dr7/algorithms/fluxcal.html}           and
  \url{https://www.sdss3.org/dr10/algorithms/magnitudes.php}       for
  further information.}.  We  compute the sky level  by averaging such
distribution in  a region  with no influence  from either  the studied
galaxy or other distant sources, where  a flat radial count profile is
displayed      (region       between      vertical       lines      in
Figure~\ref{DR7_DR10}). Then, the obtained value is subtracted from each
science frame.

To test  the accuracy of  this sky subtraction procedure,  we compared
with the  SDSS-DR10 data  release \citep[][]{ahn14} that  provides sky
subtracted and  fully calibrated frames.  In  Figure  \ref{DR7_DR10}, we
compare  the  surface-brightness  profiles and  the  $f/f_0$  $i-$band
profiles for an example galaxy (NGC~0001) using both approaches, i.e.,
our sky  subtraction scheme and  the automated procedure  performed by
the SDSS-DR10  pipeline. In  an ideal scenario,  the value  of $f/f_0$
should be  0. With our  sky subtraction  procedure we improve  the sky
level determination by a factor of  60$\%$ in the case of the $i-$band
(54.1$\%$  for $g-$band  and 53.8$\%$  for $r-$band),  allowing us  to
reach $\sim$ 1 magnitude deeper (see Figure~\ref{DR7_DR10}).

Using the sky-subtracted  images, we run {\it  ellipse} again allowing
the isophotes  to change the values  of $\epsilon$ and $PA$  to detect
changes   in  the   morphology.   These   $\epsilon$,  $PA$,   and  1D
surface-brightness profiles  along with  the previously  created masks
are then provided to the 2D photometric decomposition.

\subsection{Photometric decomposition}
\label{sec:photdec}

The structural  parameters of  the galaxy sample  were taken  from the
two-dimensional   (2D)   photometric    decomposition   described   in
\citet{mendezabreu17}.  To  this aim, we applied  the GASP2D algorithm
described  in  \citet{mendezabreu08a} and  \citet{mendezabreu14}.   We
refer  the  reader  to  these  papers for  details  about  the  actual
implementation of  the code.  In  the following we will  only describe
the specific developments introduced in this work.

\begin{figure*}
\includegraphics[angle=-90,width=\textwidth]{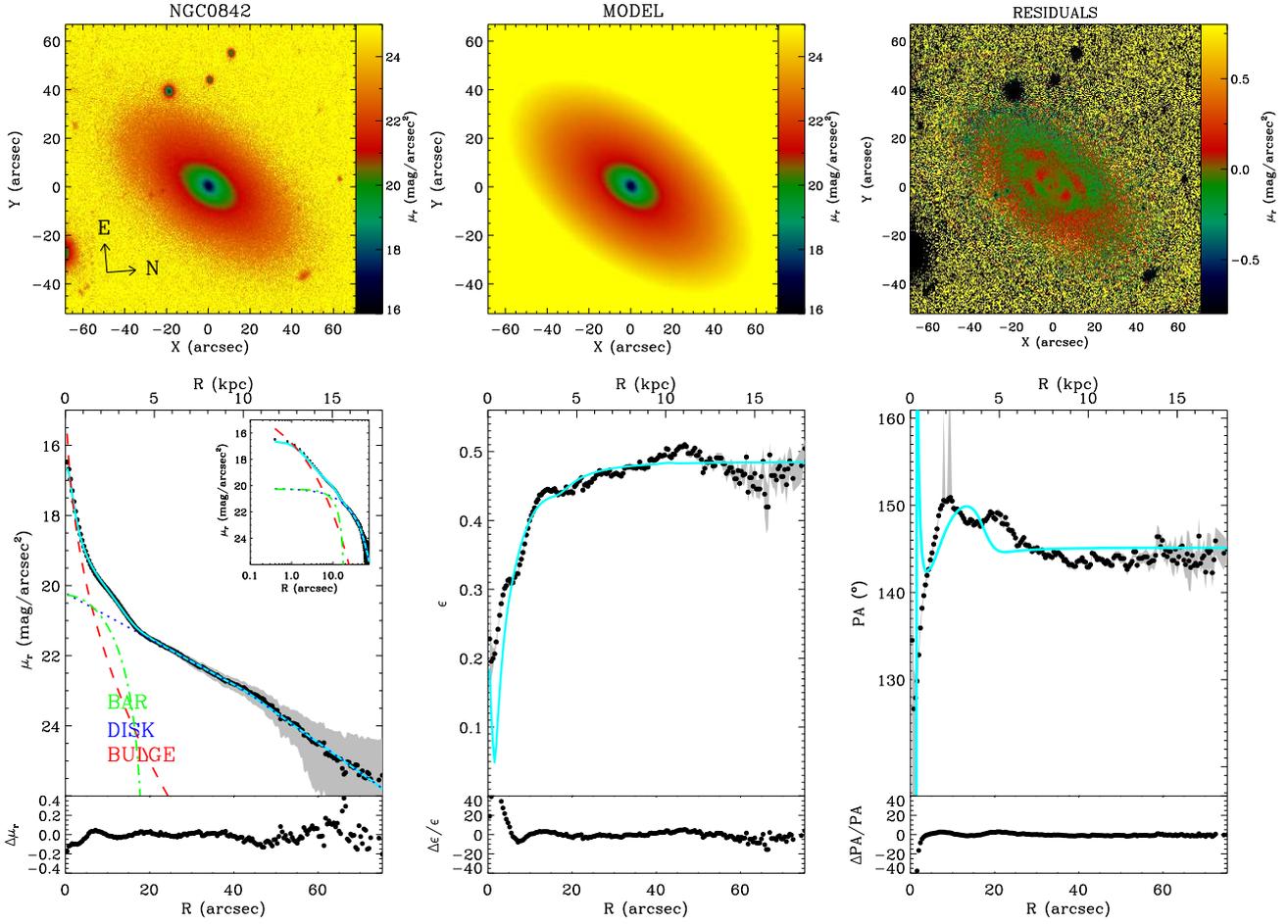}
\caption{Example of  the photometric  decomposition used  to determine
  the number of  stellar components in a galaxy.   The plot represents
  the best fit  using three components (bulge, bar, and  Type II disc)
  for the $r-$band image of  NGC~0842.  Similar plots were created for
  the  $g$- and  $i-$band.   Upper left  panel:  galaxy image.   Upper
  middle panel: best-fit model of  the galaxy image obtained by adding
  a bulge, a  bar, and a disc component.  Upper  right panel: residual
  image obtained  by subtracting  the best-fit  model from  the galaxy
  image.   Bottom  left  panel:  ellipse-averaged  surface  brightness
  radial profile of  the galaxy (black dots) and  best-fit model (cyan
  solid  line).  The  light  contributions of  the  bulge (dashed  red
  line), Type II disc (dotted blue line), and bar (dotted-dashed green
  line)   are  shown.    The  upper   inset  shows   a  zoom   of  the
  surface-brightness data  and fit  with a  logarithmic scale  for the
  distance  to  the  centre  of the  galaxy.   1D  surface  brightness
  residuals  (in  mag/arcsec$^2$  units)   are  shown  in  the  bottom
  sub-panel.  Bottom middle panel:  ellipse-averaged radial profile of
  ellipticity  of the  galaxy (black  dots) and  best-fit model  (cyan
  solid line).  1D  residuals (in percentage) are shown  in the bottom
  sub-panel.  Bottom  right panel: ellipse-averaged radial  profile of
  position angle of  the galaxy (black dots) and  best-fit model (cyan
  solid line).  1D  residuals (in percentage) are shown  in the bottom
  sub-panel. The grey shaded areas  in the bottom panels represent the
  measurement errors  derived from  the {\it  ellipse IRAF}  task when
  applied to the galaxy image.}
\label{fig:examplefit}
\end{figure*}

The  galaxy SBD  is  assumed to  be the  sum  of multiple  photometric
structures  (i.e., bulge,  disc, or  bar component)  depending on  its
specific morphology. The  inclusion of the bar SBD  in the photometric
decomposition  has been  proved to  be  critical in  order to  recover
accurate  bulge  parameters  \citep[e.g.,][]{aguerri05,laurikainen05}.
Several studies have shown that both  the S\'ersic index ($n$) and the
bulge-to-total luminosity  ratio ($B/T$) can artificially  increase if
the   bar    is   not    properly   accounted    for   in    the   fit
\citep{gadotti08,salo15}. In  addition, we allowed the  disc component
to  depart from  its {\it  purely  exponential} profile  in the  outer
regions  \citep{erwin05,pohlentrujillo06}.  Nowadays,  it is  commonly
accepted  that  galaxy discs  can  be  classified into  three  general
categories:  (i) Type  I  profiles that  follow  a single  exponential
profile along  all the optical extension  of the galaxy, (ii)  Type II
profiles that  present a  double exponential  law with  a down-bending
beyond the  so-called break radius,  and (iii) Type III  profiles that
exhibit an up-bending in the outer parts of the disc.

To account for these possibilities we adopted the following functional
parameterisation of the disc component

\begin{equation} 
I_{\rm disc} (r_{\rm disc}) = I_{\rm 0}\, \left[e^{\frac{-r_{\rm disc}}{h}}\, \theta \, + \, e^{\frac{-r_{\rm br}\,(h_{\rm out}-h)}{h_{\rm out}\,h}}\, e^{\frac{-r_{\rm disc}}{h_{\rm out}}}\,(1-\theta)\right] 
\label{eqn:disc_trunc} 
\end{equation} 
%
where
\begin{eqnarray} 
\theta  &=&   0  \qquad {\rm if} \qquad r_{\rm disc} > r_{\rm br} \nonumber \\
\theta  &=&   1  \qquad {\rm if} \qquad r_{\rm disc} < r_{\rm br}
\label{eqn:theta} 
\end{eqnarray} 
%
and $r_{\rm disc}$ is the radius measured in the Cartesian coordinates
describing  the reference  system  of the  disc.  $I_0$, $h$,  $h_{\rm
  out}$, and  $r_{\rm br}$ are  the central surface  brightness, inner
scale-length,  outer  scale-length,  and  break radius  of  the  disc,
respectively.

Figure~\ref{fig:examplefit} shows  an example  of the  photometric fit
used to  separate the  stellar structures  present in  NGC~0842. Upper
panels show the  2D SBD for the galaxy, model,  and residuals, and the
lower  panels  represent  the  1D   radial  profiles  of  the  surface
brightness, ellipticity, and position angle.  In this particular case,
the best fit is achieved using a three component model with a bulge, a
bar,  and a  Type  II  disc. The  photometric  bulge,  described by  a
S\'ersic profile, is shown with a red dashed line.

The errors on individual parameters have been computed using extensive
Monte Carlo simulations.  Mock galaxies  were generated with a variety
of  SBD  combinations  to estimate  reliable  uncertainties.   Further
details are  presented in \citet{mendezabreu17} where  the photometric
decomposition of the entire CALIFA sample is described.

Tables~\ref{tab:decompbulge},         \ref{tab:decompdisc},        and
\ref{tab:decompbar}  show the  structural parameters  derived for  our
final  sample  of {\it  bona-fide}  lenticular  galaxies described  in
Sect.~\ref{sec:S0sample}.   The  surface-brightness of  the  different
components has been corrected for both inclination using the disc axis
ratio and  Galactic extinction  \citep{schlegel98}.  No  internal dust
correction has been  attempted.

\subsection{S0 vs E separation based on the photometric decomposition}
\label{sec:S0sample}

The  initial sample  of  81 early-type  (ellipticals and  lenticulars)
galaxies  selected from  the  CALIFA sample,  and  described in  Sect.
\ref{sec:sample}, represents  the outcome of a  visual classification.
Despite its undeniable  importance, in this work we aim  to provide an
accurate  quantitative description  of  the photometric  bulges in  S0
galaxies.   Spiral  and early-type  galaxies  are  relatively easy  to
separate  based  only on  a  visual  classification, however,  a  more
thorough analysis, based on quantitative photometric decompositions is
needed  to  isolate  the  different galaxy  components  in  early-type
galaxies and to distinguish between S0 and elliptical galaxies.
The  problem  of   model  selection,  i.e.,  of   selecting  the  most
appropriate  model   that  represents  your   data  among  a   set  of
possibilities,    is    a    well-studied    topic    in    statistics
\citep[i.e.,][]{mackay03}.  In astronomy, a  clear example is provided
by  the well-known  difficulties  in separating  ellipticals from  S0s
using only photometric information.   We develop our own procedure
  to  approach  this  problem  based exclusively  on  the  photometric
  properties of  the galaxies.  The  final aim  was to obtain  a {\it
  bona-fide} sample of S0 galaxies defined in the {\it canonical way},
i.e., composed  of a   photometric  bulge dominating  the central
galaxy regions  and an outer disc  dominating the light in  the galaxy
outskirts.

The process  depicted in this section  was applied to the  81 galaxies
visually classified as either elliptical or S0, and it is based on two
steps: i) a logical filtering and ii) a statistical criteria to select
the best model.

\begin{figure}
   \centering
   \includegraphics[width=0.49\textwidth,angle=0]{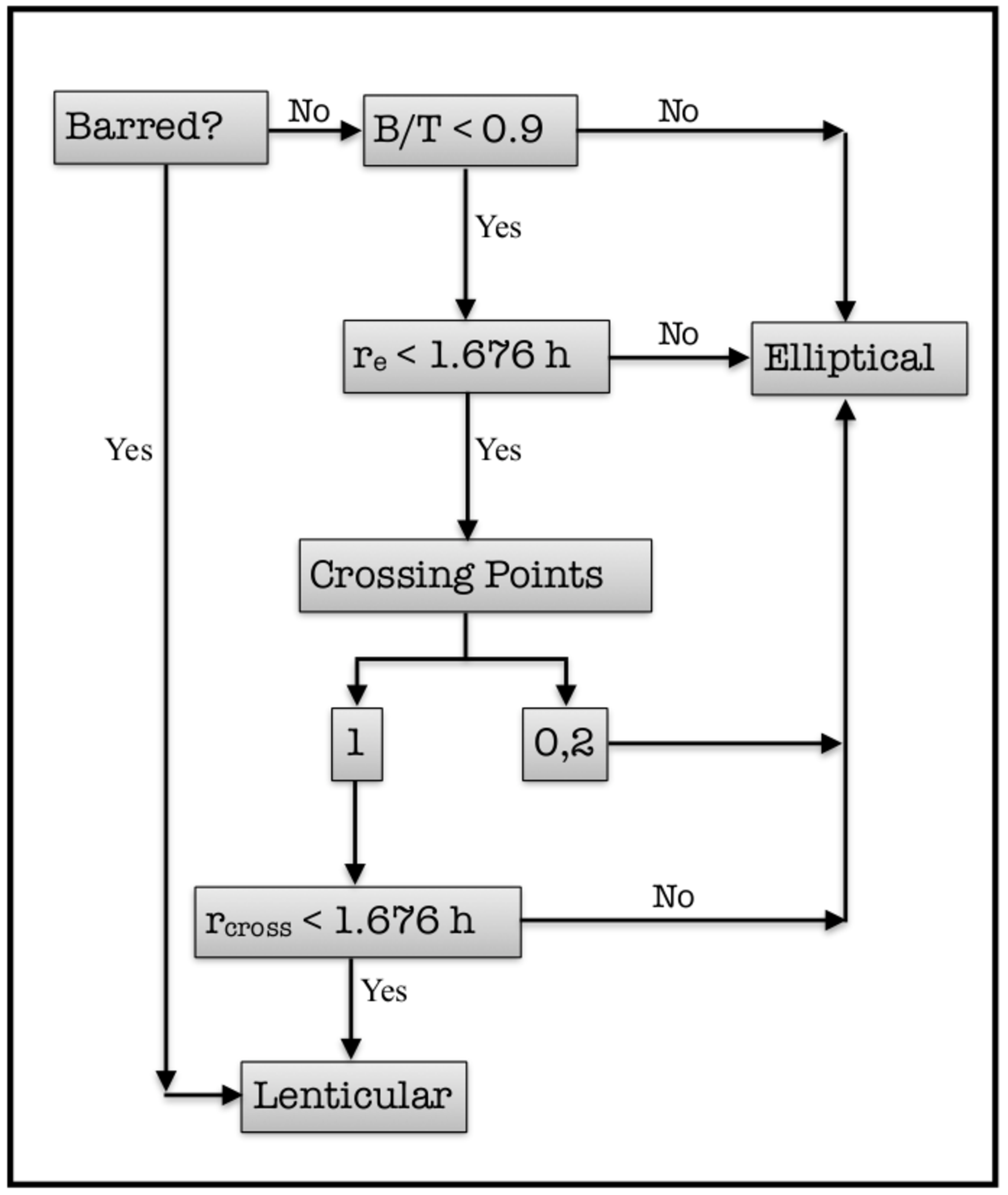}
      \caption{Logical  filter  applied  to  our  complete  sample  of
        early-type galaxies. Depending on  the answer to each question
        galaxies  were  {\it  accepted}  as  two-component  structures
        (lenticulars)  or classified  as ellipticals.   The term  {\it
          crossing point} is referred to the number of times the bulge
        profile (S\'ersic)  intersects the disc  profile (exponential)
        within the maximum  radius used for the  fit.  $r_{\rm cross}$
        indicates the radius at which this crossing point occurs.}
         \label{fig:logicalfilter}
   \end{figure}

We assume that elliptical  galaxies are photometrically well described
by  a  single  component  with   a  S\'ersic  profile.   The  simplest
description of a S0 galaxy consists  of a two-component model, i.e., a
S\'ersic  profile  describing  the  SBD  of the  bulge  and  a  single
exponential representing  the outer disc.  The  appropriateness of the
two-component model  to describe  the SBD  of our  visually classified
elliptical  and  S0 galaxies  was  evaluated  through a  {\it  logical
  filter} \citep[e.g.,][]{allen06}.   The idea behind this  step is to
provide  the  best mathematical  fit  with  a physical  meaning.   The
logical   filter   used   in   this   paper   is   shown   in   Figure
\ref{fig:logicalfilter}.   It  is  worth   noting  that  most  of  the
conditions  are set  to assure  that  the final  two-component fit  is
reliable and follow the canonical view  of S0 galaxies, i.e., an inner
dominant  bulge  with   an  outer  disc.   The   filter  will  discard
intermediate cases with embedded discs in larger elliptical galaxies.

Those galaxies  {\it accepted}  by the  logical filtering  as possibly
hosting two  components, i.e., lenticulars,  are then compared  to the
single S\'ersic fit of the whole galaxy using the Bayesian Information
Criterion \citep[$BIC$;][]{schwarz78}.  This model selection criterium
adds a penalization to the standard $\chi^2$ accounting for the number
of free parameters  included in the fit.  Thus, this  parameter can be
applied to determine  whether or not adding an  extra component (i.e.,
an outer  disc) would statistically  improve the best fit.   Under the
hypothesis of normal errors the $BIC$ statistic can be written as

\begin{equation} 
BIC = \chi^2 + k\,\ln(m) 
\label{eqn:BIC} 
\end{equation} 
%
where $k$ is  the number of free  parameters and $m$ is  the number of
independent data  points. Since in a  galaxy image not all  the pixels
are independent, we followed the prescriptions of \citet{simard11} and
substitute the number  of pixels by the number  of resolution elements
$m_{\rm res}=m/A_{\rm  psf}$ where $A_{\rm  psf}$ is the size  area of
the  Point   Spread  Function  (PSF)   at  Full  Width   Half  Maximum
(FWHM). Then Eq. \ref{eqn:BIC} can be rewritten as

\begin{equation} 
BIC = \frac{\chi^2}{A_{\rm psf}} + k\,\ln \left(\frac{m}{A_{\rm psf}}\right) 
\label{eqn:BICres} 
\end{equation} 
%

\begin{figure*}
   \centering
   \includegraphics[width=0.49\textwidth]{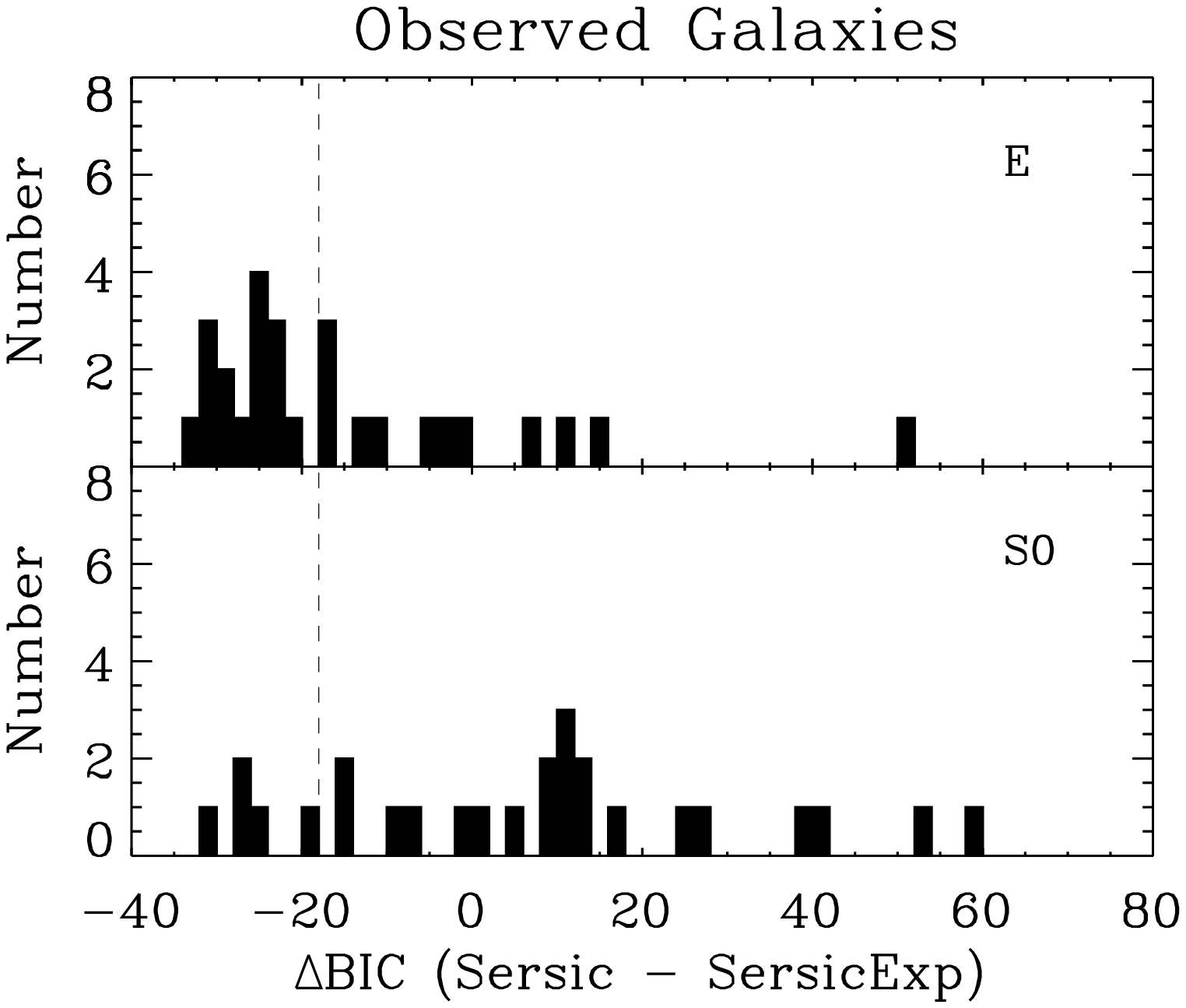}
   \includegraphics[width=0.49\textwidth]{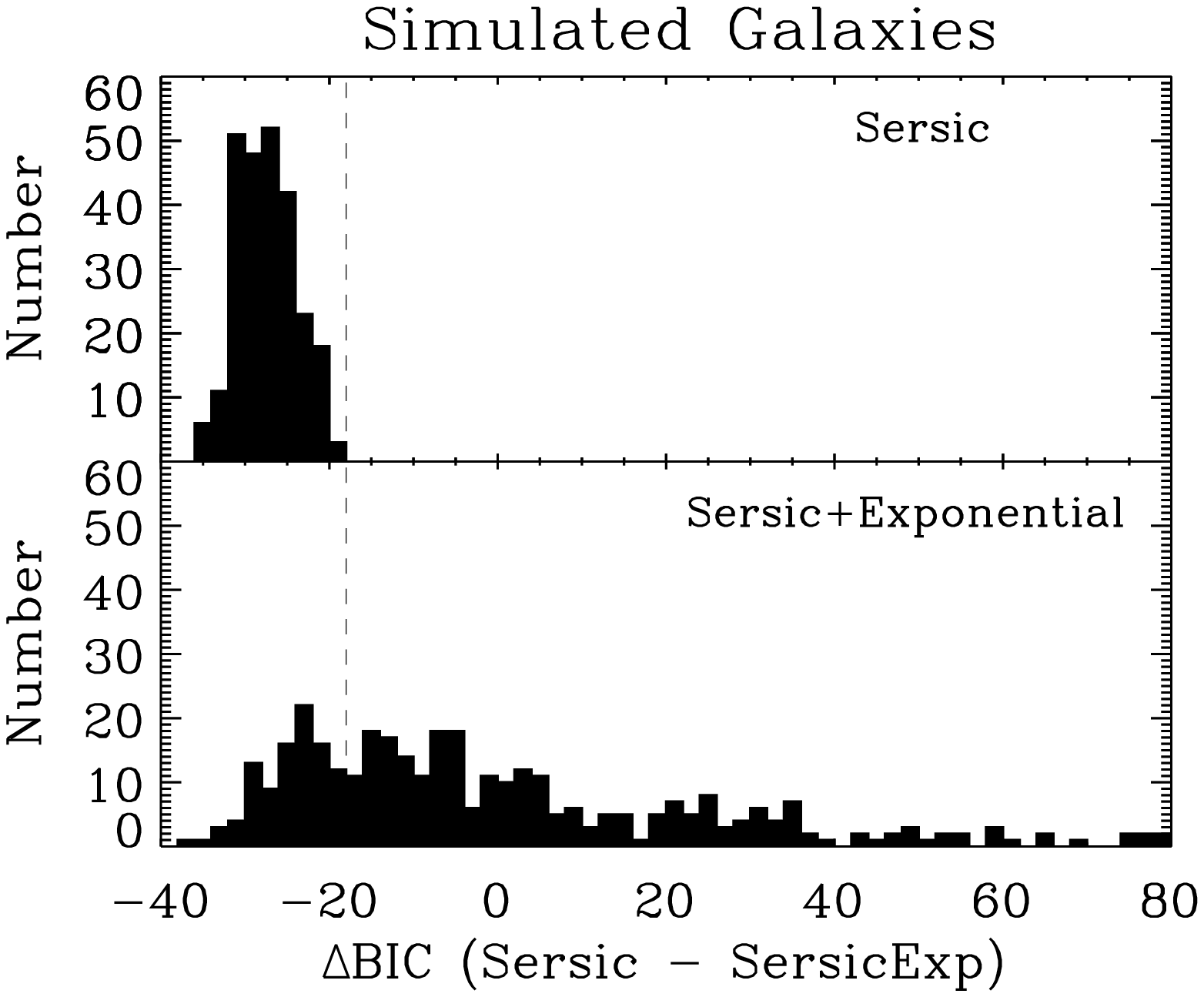}
      \caption{Distribution of $\Delta$$BIC$ values, $BIC$(S\'ersic) -
        $BIC$(S\'ersic+Exp), for our observed galaxy sample (left) and
        a  set of  simulated galaxies  (right).  The  upper and  lower
        panels  for  our  galaxy sample  represent  galaxies  visually
        classified   as   elliptical    (E)   and   lenticular   (S0),
        respectively.  The  upper and  lower panels for  the simulated
        galaxies represent the single  component S\'ersic galaxies and
        the   two   component   S\'ersic   +   Exponential   galaxies,
        respectively. The vertical dashed lines in all panels show the
        limiting $\Delta  BIC =-18$  for a galaxy  considered to  be a
        S0.}
         \label{fig:BICresult}
   \end{figure*}

Figure~\ref{fig:BICresult} (left  panels) shows the values  of $\Delta
BIC$, i.e.,  $BIC$(S\'ersic) -  $BIC$(S\'ersic+Exp), for  our visually
classified sample  of ellipticals and lenticular  galaxies that passed
the logical filter.  In this scheme, models with lower values of $BIC$
are considered  the preferred  models.  Then, $\Delta  BIC <  0$ would
imply that single component S\'ersic  models are preferred against two
components  S\'ersic +  Exponential.   Visually classified  elliptical
galaxies  are  in  good  agreement with  this  $BIC$  model  selection
criterium  except  for  4  galaxies  ($\sim$8\%).   However,  visually
classified lenticular  galaxies span  a wider  range of  $BIC$ values.
The actual line of demarcation for  strong evidence against one of the
models is  however not clear.  Some  studies have proposed a  value of
$\Delta BIC  > 10$ as  division for  a very strong  preference against
higher $BIC$ models  \citep{kass95}, but in complex cases  such as the
one presented here  a {\it calibration} of the  $\Delta BIC$ parameter
using mock galaxy simulations is preferred.

Mock galaxies were created  as in Sect. \ref{sec:photdec} (photometric
error computation) and therefore they provide a good representation of
the real galaxies  with the same observational SDSS setup.   We used a
sample of  $\sim$250 single S\'ersic component  galaxies and $\sim$350
two  component S\'ersic  +  Exponential galaxies.   Both samples  were
fitted  again as  if  they  were real  galaxies  using  both a  single
S\'ersic component and  a two component S\'ersic  + Exponential model,
and  the  $BIC$   statistics  was  computed  as   for  real  galaxies.
Figure~\ref{fig:BICresult} (right  panels) shows the  results obtained
for  the  simulated  mock  galaxies.    As  for  real  galaxies,  mock
elliptical galaxies  show a  narrow distribution  of the  $\Delta BIC$
statistics  with  all  galaxies  showing  $\Delta  BIC  <  -18$.   The
distribution  of  lenticular galaxies  is  also  similar to  the  real
galaxies, strongly overlapping with the region defined by ellipticals.
These  results  highlight  the intrinsic  difficulties  of  separating
ellipticals  from S0  galaxies  using photometric  data,  but it  also
provides us  with a method  to define  {\it bona-fide} S0  galaxies as
those with $\Delta BIC > -18$,  since no ellipticals lie in this $BIC$
range  of values.   It is  worth noting  that another  model selection
statistics    such     as    the    Akaike     Information    Criteria
\citep[$AIC$;][]{akaike74}  was  also  used in  this  study  obtaining
similar  results.  Nevertheless,  the  $AIC$ penalizes  the number  of
parameters less strongly than the $BIC$ does and therefore we restrict
our  further analysis  to the  $BIC$ selected  sample to  minimise the
number of {\it false-positive} detections due to overfitting.

\begin{figure*}
   \centering
   \includegraphics[bb=84 60 620 594]{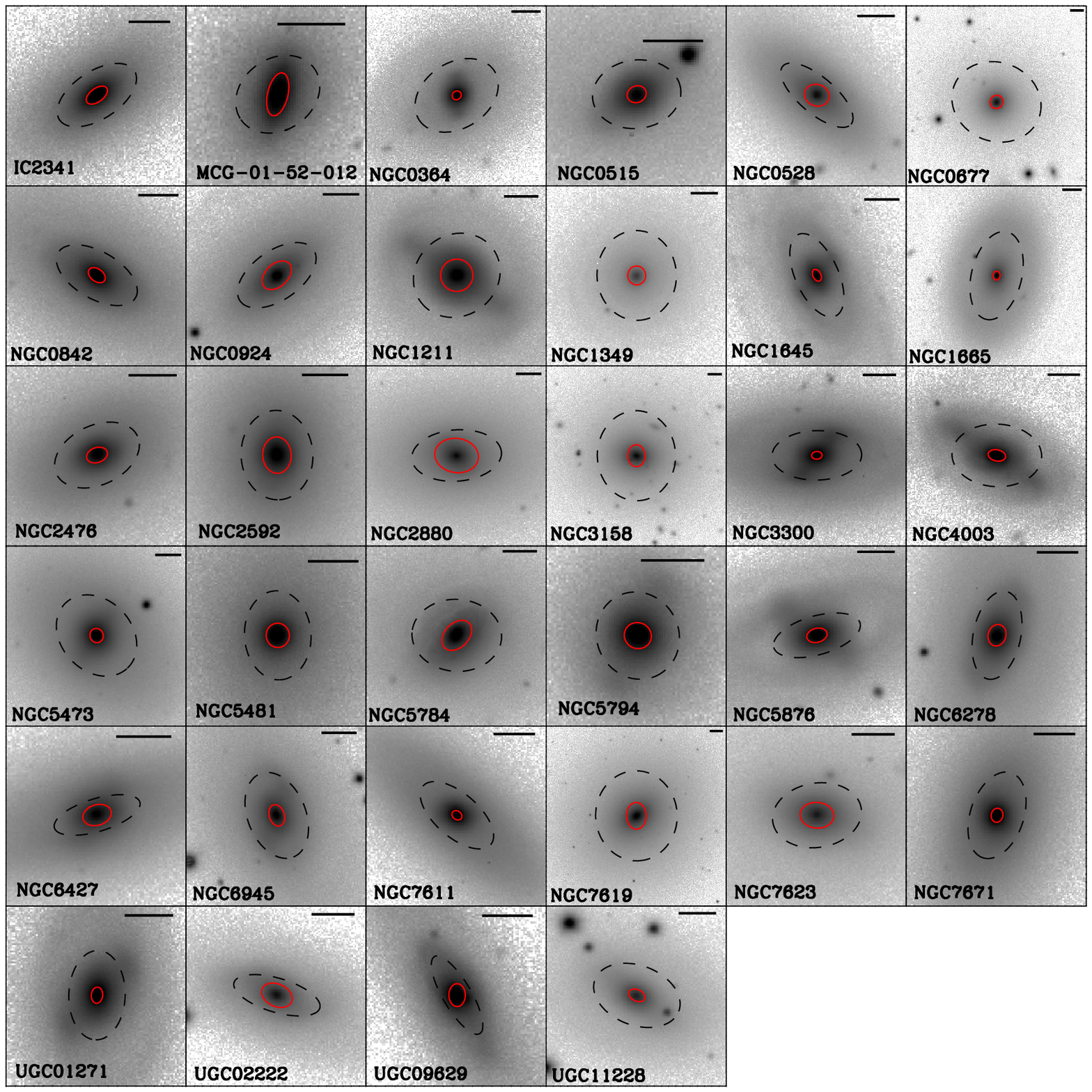}
      \caption{SDSS  $r-$band  images of  the  34  CALIFA S0  galaxies
        presented in this  study. In all images, north is  up and east
        is left.  The black dashed  ellipses show the galaxy effective
        radius   as   computed   from    the   light   growth   curves
        \citep[see][]{walcher14}.   The red  solid  ellipse shows  the
        effective  radius  and  geometry  of  the  photometric  bulges
        obtained in this paper. The upper bar in each panel represents
        10 arcsecs.}
         \label{fig:mosaic}
   \end{figure*}

Summarizing, all galaxies with additional structural components (i.e.,
bars  or non-single  exponential  discs) were  directly classified  as
lenticular galaxies.  For the  remaining galaxies, those classified by
the logical  filter as elliptical and  with a $\Delta BIC  < -18$ were
photometrically classified  as ellipticals.  On the  other hand, those
accepted by the logical filter as two-component and with $\Delta BIC >
-18$  represent  our  final  sample  of  {\it  bona-fide}  photometric
lenticular galaxies.  Finally, those  galaxies accepted by the logical
filter as two-component  and with $\Delta BIC < -18$  cannot be safely
classified  and they  were labelled  in our  sample as  {\it unknown}.
This latter  group has not been  used in any further  analysis in this
paper. There are not galaxies  classified as elliptical by the logical
filter and with $\Delta BIC > -18$.  Table \ref{tab:ellipvss0} shows a
compendium of the number of galaxies  in each sample. The final sample
studied   in   this   work   is    composed   by   34   S0   galaxies.
Figure~\ref{fig:mosaic} shows  a mosaic with the  thumbnail images for
our  S0 galaxies.   The effective  radius of  the whole  galaxy (black
dashed), as computed from  the growth curves \citep[see][]{walcher14},
is shown and the image size  has been rescaled accordingly.  The bulge
effective radius  ($r_{e}$, red)  is also  shown.  This  parameter was
obtained from the photometric decomposition described in this section.

\begin{table}
\caption{Schematic  of  the  sample selection  process.  E-Elliptical,
  S0-Lenticular, U-Unknown.   (1) Number of galaxies  using the CALIFA
  visual  classification; (2)  number  of galaxies  after the  logical
  filtering; (3)  number of galaxies  after the logical  filtering and
  $BIC$ classification; (4) final sample used in this study. }
\begin{center}
\begin{tabular}{|c|c|c|c|}
\hline
CALIFA VISUAL       &  LF   & LF + $\Delta BIC$   &  FINAL                \\
  (1)               &  (2)  &     (3)             &   (4)                 \\
\hline
    \multirow{2}{*}{48 E}& 21 E                   & 21 E     & \\ \cline{2-3}
                         & \multirow{2}{*}{27 S0} & 15 U     & \\
                         &                        & 12 S0    & 26 E \\
    \cline{1-3}
    \multirow{2}{*}{33 S0}& 5 E                   & 5E       & 21 U\\\cline{2-3}
                          & \multirow{2}{*}{28 S0}& 6 U      & 34 S0\\
                          &                       & 22 S0      & \\
    \hline
\end{tabular}

\end{center}
\label{tab:ellipvss0}
\end{table}

\section{Global properties of the galaxy sample}
\label{sec:globalprop}

\begin{figure*}
   \centering
   \includegraphics[bb= 54 520 558 720]{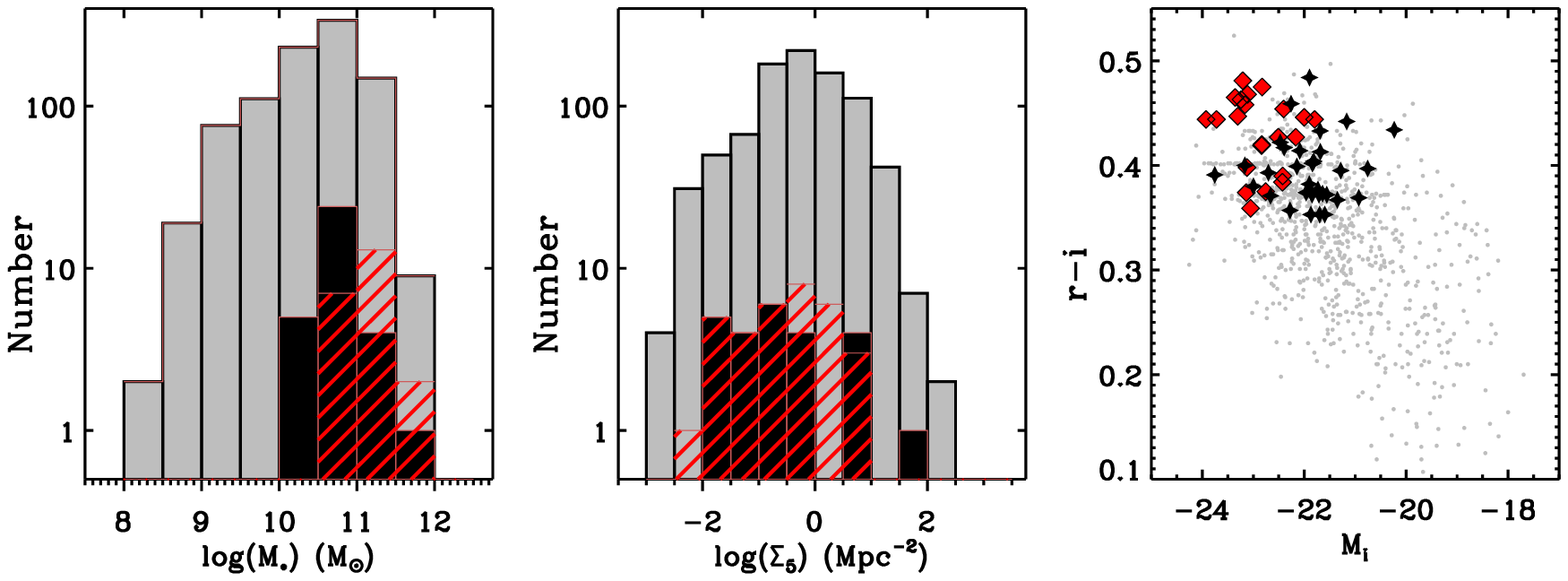}
      \caption{{\it Left  panel}.  Distribution of the  galaxy stellar
        masses.   {\it Middle  panel}.  Distribution  of local  galaxy
        densities.  {\it Right  panel}.  Colour-magnitude  diagram.  In
        all  panels  grey colours  represent  the  whole CALIFA  mother
        sample, black  and red  show the sample  of S0  and elliptical
        galaxies selected in this  work, respectively.  The quantities
        shown    in   the    panels   have    been   extracted    from
        \citet{walcher14}.}
         \label{fig:galprop}
   \end{figure*}

Figure  \ref{fig:galprop} shows  the  range of  stellar masses,  local
galaxy  densities,  and  colours  probed  by our  final  sample  of  S0
galaxies. For  comparison, we  have also included  the values  for the
CALIFA mother sample  and the elliptical sample that will  be used for
comparison in Sect~\ref{sec:photkin}.

From Figure  \ref{fig:galprop} (left  panel) it is  clear that  our S0
sample covers a narrow range  of stellar masses, $M_{\star}/M_{\sun} >
10^{10}$. Compared to  the elliptical sample they  show slightly lower
masses but they  both represent the high mass end  of the whole CALIFA
mother sample \citep[e.g.,][]{gonzalezdelgado15}.   Although the tight
stellar mass range covered by our  sample is not representative of the
wide  range  of masses  encompassed  by  the  whole population  of  S0
galaxies,  it  allows us  to  characterise  a well-defined  sample  of
high-mass S0 galaxies.

The  environment where  our S0  galaxies live  is presented  in Figure
\ref{fig:galprop}  (middle panel).   The local  galaxy densities  were
extracted  from  \citet{walcher14}  and   they  were  computed  as  in
\citet{aguerri09}.  Despite S0 galaxies being found in a wide range of
local densities, our  sample is mainly composed of  galaxies living in
low-density environments  ($\Sigma_5 < 1 $  gal/Mpc$^{2}$). Therefore,
we are not probing S0 in galaxy clusters. We have further checked this
by studying the  membership of our S0 sample  within well-known galaxy
structures.   We  found  that  none   of  our  galaxies  belong  to  a
high-density  structure  \citep[see][for  details  on  the  membership
  definition]{walcher14}.

The S0 galaxy colours shown  in Figure \ref{fig:galprop} (right panel)
show that they lie on the  red-sequence.  Their colours are similar to
the reddest galaxies  in the CALIFA mother sample  and comparable with
those of the elliptical galaxies.

To summarise, our galaxy sample  represents a well characterised sample
of high mass, red, and relatively isolated S0 galaxies.

\section{Stellar kinematic measurements}
\label{sec:kinmeasurements}

\subsection{Stellar kinematics maps}
\label{sec:kinematics}

The stellar  kinematics of  the galaxy sample  were measured  from the
spectral  datacubes  observed with  the  V1200  grating. An  extensive
description     of     the     methodology     is     explained     in
\cite{falconbarroso17}.  We  briefly describe  in the  following the
main characteristics of the process.

The spaxels of  the datacube were binned using  a Voronoi tessellation
method \citep{cappellaricopin03} in order to achieve a signal-to-noise
(S/N) ratio  $>$ 20.   Spaxels with  S/N $<$ 3  were removed  from the
analysis.   The  first  two  moments  of  the  line-of-sight  velocity
distribution (LOSVD) were then measured for each Voronoi bin using the
penalised        pixel-fitting        method        (pPXF)        from
\citet{cappellariemsellem04}. The possibility  of fitting higher-order
moments of the LOSVD was turned off  during the fit due to the limited
S/N of the  spectra. A non-negative linear combination of  a subset of
328 stellar  templates from  the Indo-US library  \citep{valdes04} was
used  to fit  the  spectra.  The  final errors  in  both velocity  and
velocity dispersion were obtained via Monte Carlo simulations.

\subsection{Stellar kinematic properties}
\label{sec:kinprop}

Historically,  most of  the studies  in galaxy  bulge kinematics  were
based on long-slit spectroscopy, therefore most of these studies focus
on  edge-on disc  galaxies  in  order to  avoid  as  much as  possible
contamination from  the disc  component.  Then,  slits were  placed at
different heights over the disc  plane to compute the maximum rotation
velocity. With the advent of IFS spectroscopy, a re-formulation of the
$v/\sigma$ vs  $\epsilon$ diagram  was done by  \citet{binney05}.  The
updated  formulae   to  compute  the  $v/\sigma$   relation  using  2D
spectroscopy can be defined as

\begin{equation} 
\left( \frac{v}{\sigma} \right)^2_{R} = \frac{\langle v^2 \rangle}{\langle \sigma^2 \rangle} = \frac{\sum_i^N F_i v^2_i}{\sum_i^N F_i \sigma^2_i}
\label{eqn:vsigma} 
\end{equation} 
%
where $F_i$  is the flux  contained inside  the $i$th Voronoi  bin and
$v_i$ and $\sigma_i$ are the  corresponding measured mean velocity and
velocity dispersion.

According to  this new formulation,  and in  their quest for  a better
representation    of    the    dynamical    support    of    galaxies,
\citet{emsellem07} defined a new  kinematic parameter, $\lambda$, as a
function of surface brightness weighted  averages of $v$ and $\sigma$.
Furthermore, they  included a  factor depending on  the galactocentric
distance in order  to capture the spatial information  provided by the
IFS, thus converting  $\lambda$ into a proxy for  the specific angular
momentum. The equation to measure $\lambda$ takes the form

\begin{equation} 
\lambda_R = \frac{\sum_i^N F_i\,R_i\,|v_i|}{\sum_i^N F_i\,R_i\,\sqrt{v^2_i + \sigma^2_i}}
\label{eqn:lambda} 
\end{equation} 
%
where $F_i$ is the flux inside the $i$th bin, $R_i$ is the distance to
the galaxy  centre, and  $v_i$ and  $\sigma_i$ the  corresponding mean
stellar velocity and velocity dispersion.

$\lambda_R$  is by  definition  a  function of  the  radius, thus  its
integrated value will depend on the  radial extension over which it is
measured. Previous works  carried out by the SAURON  and ATLAS3D teams
\citep{emsellem07, emsellem11} have used the half light radius ($r_e$)
of  the whole  galaxy.  This  quantity is  relatively easy  to measure
(using the curve of growth obtained from ellipse fitting to the galaxy
isophotes) and  provides a  single parameterisation of  the rotational
support of  the galaxy independently  of morphology. However,  in this
work we  are interested  in the  kinematics of  the galaxy  bulges and
therefore we computed the values of both $v/\sigma$ and $\lambda$ over
1   effective  radius   of  the   photometric  bulge   component  (see
Sect~\ref{sec:photdec}).  A  complete analysis  on the  $v/\sigma$ and
$\lambda$  properties of  the whole  galaxy, and  the comparison  with
previous surveys, will be given in Falcon-Barroso et al. (in prep.)

The final  errors in our integrated  kinematic properties ($v/\sigma$,
and $\lambda$) come from three main sources: the measurement errors of
the  stellar   kinematic  maps   \citep[see][]{falconbarroso17},  the
effects of  pixelization and PSF associated  with measuring integrated
properties in  small apertures,  and the  errors corresponding  to the
correction  for  the  disc  kinematics.   All  errors  were  added  in
quadrature.  The measurement errors  were propagated to the integrated
quantities  by  using Monte  Carlo  simulations  of the  velocity  and
velocity dispersion  maps, i.e., varying  randomly the values  in each
spaxel  within  their error.   The  pixelation  and PSF  effects  were
estimated   using  mock   datacube  spectroscopic   simulations.   The
methodology  is explained  in detail  in Appendix~\ref{sec:psf}.   The
impact  of the  disc kinematics  in  the bulge  measurements are  also
estimated  using  mock  spectroscopic   simulations  as  described  in
Sect.~\ref{sec:disccontamination}  and compared with  the results
  from  Schwarzschild   dynamical  modelling  of  the   galaxies  (see
  Sect.~\ref{sec:schwarzschild}).   The  final  corrected  values  of
$\lambda$  and  $v/\sigma$,  their edge-on  deprojections,  and  their
corresponding errors are shown in Table~\ref{tab:kinematics}.

\begin{table*}
\caption{Kinematic values measured  for our sample of  34 S0 galaxies.
  (1) Galaxy name; (2) $\lambda_{e,b}$  measured within 1 $r_{e,b}$ of
  the bulge  (m); (3) $\lambda_{e,b}$  measured within 1  $r_{e,b}$ of
  the bulge,  corrected for  pixelation and resolution  effects (p+r);
  (4)  $\lambda_{e,b}$  measured  within  1 $r_{e,b}$  of  the  bulge,
  corrected for pixelation, resolution,  and disc contamination (these
  values  are   used  throughout   the  paper,  p+r+d);   (5)  edge-on
  deprojected value  of $\lambda_{e,b}$ (p+r+d);  (6) $v/\sigma_{e,b}$
  measured within 1  $r_{e,b}$ of the bulge  (m); (7) $v/\sigma_{e,b}$
  measured within 1  $r_{e,b}$ of the bulge,  corrected for pixelation
  and resolution effect (p+r);  (8) $v/\sigma_{e,b}$ measured within 1
  $r_{e,b}$ of  the bulge,  corrected for pixelation,  resolution, and
  disc  contamination (these  values  are used  throughout the  paper,
  p+r+d); (9)  edge-on deprojected value of  $v/\sigma_{e,b}$ (p+r+d);
  (10) intrinsic ellipticity of the  bulge obtained assuming that both
  the bulge and the disc are oblate ellipsoids.}
\begin{center}
\resizebox{2.1\columnwidth}{!}{%
\begin{tabular}{cccccccccc}
\hline
Galaxy & $\lambda_{e,b}$ & $\lambda_{e,b}$ & $\lambda_{e,b}$ & $\lambda_{e,b,0}$ & $v/\sigma_{e,b}$  & $v/\sigma_{e,b}$  & $v/\sigma_{e,b}$  & $v/\sigma_{e,b,0}$ & $\epsilon_{intr,e,b}$  \\
       &        (m)            &        (p+r)               &         (p+r+d)               &  (edge-on) &         (m)             &       (p+r)                  &      (p+r+d)                   & (edge-on) &     \\
  (1)  &      (2)           &   (3)                 &     (4)                &  (5)                 &     (6)                 &      (7)                &   (8)       &   (9)   &  (10) \\
\hline
IC2341        & 0.30$\pm$0.04 & 0.44$\pm$0.04 & 0.38$\pm$0.07 & 0.41$\pm$0.07  &  0.30$\pm$0.04  &    0.42$\pm$0.04  &    0.33$\pm$0.06  & 0.35$\pm$0.07 & 0.69      \\
MCG-01-52-012 & 0.22$\pm$0.02 & 0.32$\pm$0.02 & 0.29$\pm$0.06 & 0.37$\pm$0.10  &  0.21$\pm$0.02  &    0.29$\pm$0.02  &    0.25$\pm$0.03  & 0.33$\pm$0.11 & 0.28      \\
NGC0364       & 0.19$\pm$0.04 & 0.34$\pm$0.04 & 0.33$\pm$0.04 & 0.38$\pm$0.09  &  0.20$\pm$0.03  &    0.32$\pm$0.03  &    0.31$\pm$0.03  & 0.38$\pm$0.10 & 0.26   \\
NGC0515       & 0.12$\pm$0.05 & 0.22$\pm$0.05 & 0.21$\pm$0.06 & 0.27$\pm$0.10  &  0.13$\pm$0.04  &    0.26$\pm$0.04  &    0.20$\pm$0.04  & 0.26$\pm$0.10 & 0.44       \\
NGC0528       & 0.26$\pm$0.07 & 0.46$\pm$0.07 & 0.39$\pm$0.07 & 0.39$\pm$0.07  &  0.28$\pm$0.07  &    0.41$\pm$0.07  &    0.34$\pm$0.07  & 0.34$\pm$0.10 & 0.19   \\
NGC0677       & 0.10$\pm$0.01 & 0.14$\pm$0.01 & -             & -              &  0.11$\pm$0.01  &    0.14$\pm$0.01  &    -              & -             & 0.24        \\
NGC0842       & 0.26$\pm$0.05 & 0.44$\pm$0.05 & 0.42$\pm$0.05 & 0.44$\pm$0.07  &  0.26$\pm$0.04  &    0.40$\pm$0.04  &    0.38$\pm$0.04  & 0.41$\pm$0.07 & 0.50    \\
NGC0924       & 0.39$\pm$0.05 & 0.62$\pm$0.05 & 0.59$\pm$0.05 & 0.61$\pm$0.08  &  0.42$\pm$0.05  &    0.59$\pm$0.05  &    0.55$\pm$0.06  & 0.60$\pm$0.10 & 0.52    \\
NGC1211       & 0.19$\pm$0.01 & 0.27$\pm$0.01 & -             & -              &  0.20$\pm$0.01  &    0.26$\pm$0.01  &    -              & -             & 0.35    \\
NGC1349       & 0.12$\pm$0.01 & 0.19$\pm$0.02 & 0.16$\pm$0.05 & 0.25$\pm$0.15  &  0.13$\pm$0.02  &    0.19$\pm$0.02  &    0.17$\pm$0.02  & 0.27$\pm$0.12 & 0.23    \\
NGC1645       & 0.20$\pm$0.06 & 0.35$\pm$0.06 & 0.33$\pm$0.06 & 0.34$\pm$0.08  &  0.21$\pm$0.06  &    0.34$\pm$0.06  &    0.32$\pm$0.06  & 0.33$\pm$0.08 & 0.48    \\
NGC1665       & 0.12$\pm$0.01 & 0.23$\pm$0.01 & 0.22$\pm$0.01 & 0.23$\pm$0.03  &  0.13$\pm$0.01  &    0.21$\pm$0.01  &    0.20$\pm$0.01  & 0.22$\pm$0.03 & 0.28     \\ 
NGC2476       & 0.19$\pm$0.06 & 0.33$\pm$0.06 & 0.30$\pm$0.06 & 0.32$\pm$0.11  &  0.20$\pm$0.06  &    0.31$\pm$0.06  &    0.28$\pm$0.06  & 0.32$\pm$0.11 & 0.57     \\
NGC2592       & 0.32$\pm$0.02 & 0.49$\pm$0.02 & 0.45$\pm$0.05 & 0.54$\pm$0.12  &  0.31$\pm$0.02  &    0.44$\pm$0.02  &    0.36$\pm$0.04  & 0.48$\pm$0.14 & 0.75     \\
NGC2880       & 0.37$\pm$0.02 & 0.49$\pm$0.02 & -             & -              &  0.38$\pm$0.02  &    0.46$\pm$0.02  &    -              & -             & 0.30     \\ 
NGC3158       & 0.21$\pm$0.01 & 0.28$\pm$0.01 & -             & -              &  0.25$\pm$0.01  &    0.30$\pm$0.01  &    -              & -             & 0.76         \\
NGC3300       & 0.11$\pm$0.05 & 0.19$\pm$0.05 & 0.18$\pm$0.12 & 0.20$\pm$0.07  &  0.11$\pm$0.05  &    0.18$\pm$0.05  &    0.17$\pm$0.14  & 0.19$\pm$0.06 & 0.36     \\
NGC4003       & 0.30$\pm$0.06 & 0.48$\pm$0.06 & 0.44$\pm$0.05 & 0.49$\pm$0.11  &  0.31$\pm$0.07  &    0.45$\pm$0.07  &    0.39$\pm$0.05  & 0.46$\pm$0.16 & 0.76     \\
NGC5473       & 0.12$\pm$0.03 & 0.18$\pm$0.03 & 0.14$\pm$0.06 & 0.19$\pm$0.11  &  0.12$\pm$0.03  &    0.18$\pm$0.03  &    0.14$\pm$0.08  & 0.19$\pm$0.12 & 0.20     \\
NGC5481       & 0.09$\pm$0.02 & 0.17$\pm$0.02 & 0.14$\pm$0.03 & 0.18$\pm$0.07  &  0.10$\pm$0.02  &    0.15$\pm$0.02  &    0.13$\pm$0.03  & 0.16$\pm$0.07 & 0.15         \\
NGC5784       & 0.22$\pm$0.02 & 0.32$\pm$0.02 & -             & -              &  0.23$\pm$0.02  &    0.30$\pm$0.02  &    -              & -             & 0.64         \\
NGC5794       & 0.17$\pm$0.02 & 0.30$\pm$0.02 & 0.27$\pm$0.05 & 0.42$\pm$0.17  &  0.18$\pm$0.02  &    0.28$\pm$0.02  &    0.26$\pm$0.03  & 0.46$\pm$0.22 & 0.11         \\
NGC5876       & 0.17$\pm$0.04 & 0.25$\pm$0.04 & 0.24$\pm$0.04 & 0.25$\pm$0.05  &  0.17$\pm$0.04  &    0.24$\pm$0.04  &    0.23$\pm$0.04  & 0.23$\pm$0.04 & 0.38      \\
NGC6278       & 0.17$\pm$0.03 & 0.27$\pm$0.03 & 0.26$\pm$0.03 & 0.28$\pm$0.05  &  0.17$\pm$0.03  &    0.24$\pm$0.03  &    0.24$\pm$0.03  & 0.25$\pm$0.05 & 0.25      \\
NGC6427       & 0.24$\pm$0.05 & 0.38$\pm$0.05 & 0.36$\pm$0.10 & 0.36$\pm$0.05  &  0.24$\pm$0.04  &    0.35$\pm$0.04  &    0.34$\pm$0.13  & 0.34$\pm$0.04 & 0.32       \\
NGC6945       & 0.17$\pm$0.03 & 0.29$\pm$0.03 & 0.28$\pm$0.03 & 0.31$\pm$0.06  &  0.18$\pm$0.03  &    0.26$\pm$0.03  &    0.25$\pm$0.03  & 0.29$\pm$0.06 & 0.60       \\
NGC7611       & 0.13$\pm$0.05 & 0.25$\pm$0.05 & 0.25$\pm$0.04 & 0.26$\pm$0.06  &  0.15$\pm$0.03  &    0.27$\pm$0.03  &    0.26$\pm$0.04  & 0.28$\pm$0.04 & 0.29       \\
NGC7619       & 0.11$\pm$0.01 & 0.13$\pm$0.01 & -             & -              &  0.12$\pm$0.01  &    0.14$\pm$0.01  &    -              & -             & 0.71           \\
NGC7623       & 0.14$\pm$0.02 & 0.21$\pm$0.02 & 0.16$\pm$0.05 & 0.20$\pm$0.10  &  0.14$\pm$0.01  &    0.19$\pm$0.01  &    0.17$\pm$0.04  & 0.21$\pm$0.07 & 0.58       \\
NGC7671       & 0.16$\pm$0.02 & 0.31$\pm$0.02 & 0.30$\pm$0.02 & 0.33$\pm$0.05  &  0.16$\pm$0.03  &    0.27$\pm$0.03  &    0.26$\pm$0.02  & 0.29$\pm$0.06 & 0.26       \\
UGC01271      & 0.20$\pm$0.03 & 0.35$\pm$0.03 & 0.34$\pm$0.03 & 0.37$\pm$0.07  &  0.20$\pm$0.03  &    0.32$\pm$0.04  &    0.31$\pm$0.04  & 0.35$\pm$0.07 & 0.46        \\
UGC02222      & 0.28$\pm$0.03 & 0.44$\pm$0.03 & 0.40$\pm$0.04 & 0.40$\pm$0.04  &  0.29$\pm$0.04  &    0.41$\pm$0.04  &    0.40$\pm$0.04  & 0.40$\pm$0.04 & 0.36        \\
UGC09629      & 0.21$\pm$0.03 & 0.36$\pm$0.04 & 0.35$\pm$0.04 & 0.36$\pm$0.05  &  0.21$\pm$0.04  &    0.32$\pm$0.04  &    0.31$\pm$0.04  & 0.32$\pm$0.05 & 0.28            \\
UGC11228      & 0.12$\pm$0.02 & 0.20$\pm$0.02 & 0.18$\pm$0.04 & 0.21$\pm$0.05  &  0.12$\pm$0.02  &    0.18$\pm$0.02  &    0.17$\pm$0.04  & 0.20$\pm$0.05 & 0.79        \\
\hline
\end{tabular}%
} 

\end{center}
\label{tab:kinematics}
\end{table*}                            

\section{Results}
\label{sec:results}

\subsection{Structural components and photometric properties of the sample}
\label{sec:photprop}

In this  section we dissect  the structural components present  in our
sample of 34 S0 galaxies.  A comparison with previous results from the
literature using  similar methodologies, but larger  samples, allow us
to place our photometric components in a more general context.

\subsubsection{Bulge properties}

\begin{figure*}
   \centering
   \includegraphics[bb= 54 420 558 620]{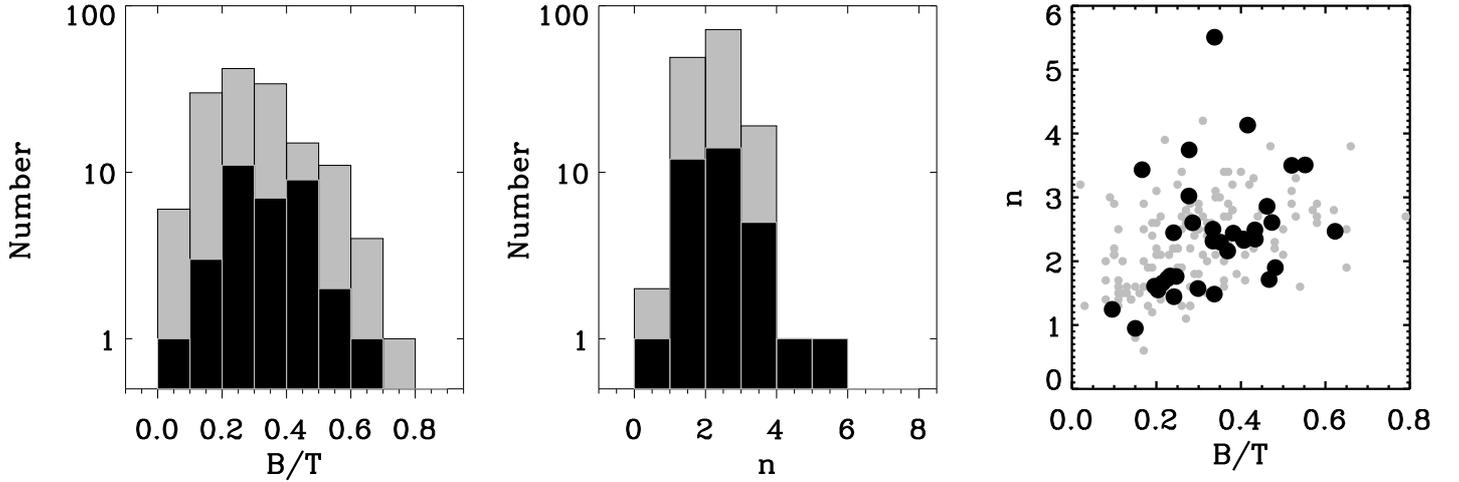}
      \caption{{\it  Left panel.}   Distribution  of $B/T$  luminosity
        ratio.  {\it Middle panel.}  S\'ersic index ($n$).  {\it Right
          panel.}   $B/T$   vs  $n$   (right  panel).    Black  filled
        histograms represent our S0  sample.  Grey histograms show the
        results from  the S0  sample of  \citet{laurikainen10}.  Black
        and grey points  represent our S0 sample and  the results from
        \citet{laurikainen10}, respectively.}
         \label{fig:bulphot}
   \end{figure*}

Figure~\ref{fig:bulphot} shows the $i-$band  distribution of the $B/T$
luminosity ratio, S\'ersic index, and their correlation for the bulges
of our  galaxy sample.   The $B/T$ distribution  is compared  with the
sample of S0  galaxies from \citet{laurikainen10} which  uses the same
photometric  definitions for  the  different  galaxy components.   The
distributions are  in good  agreement showing a  wide range  of values
from small bulges ($B/T \sim$ 0.1) to galaxies with large bulges ($B/T
\sim 0.6$). The  S\'ersic index distribution also shows  a large range
of values and a similar distribution to that of \citet{laurikainen10}.
These  two  parameters  are  commonly  used  to  describe  bulges  and
occasionally they are  used interchangeably.  Figure~\ref{fig:bulphot}
shows the  correlation between $B/T$  and $n$.  Despite the  fact that
high $n$ bulges  show larger values of $B/T$, the  correlation is weak
(Pearson coefficient  $\rho \sim 0.5$)  and there is large  scatter in
the  relation, with  highly concentrated  bulges ($n  \geq 3$)  can be
found in galaxies with either large or small $B/T$ ratios.

\begin{figure}
   \centering
   \includegraphics[width=0.49\textwidth]{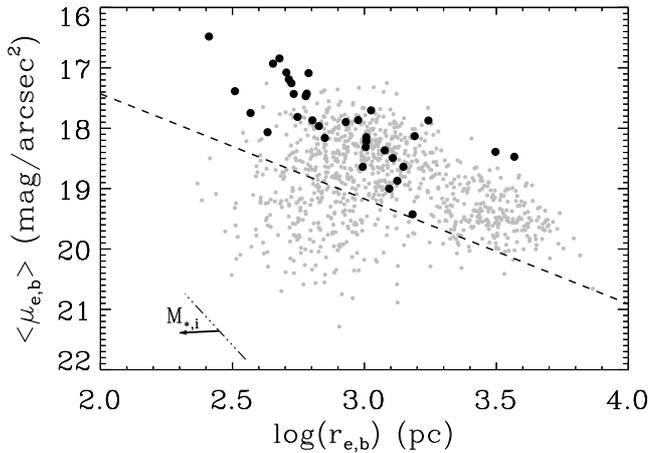}
      \caption{Mean effective surface  brightness within the effective
        radius ($\mu_{e,b}$) vs the  logarithm of the effective radius
        ($r_{e,b}$) for  the S0 bulges  in our sample  (black points).
        Small grey points show  the results from \citet{gadotti09} for
        a galaxy sample including spiral and elliptical galaxies.  The
        dotted  line represents  the  line  dividing classical  bulges
        (above)   from   disc-like   bulges  (below)   following   the
        prescription of \citet{gadotti09}.  The dash-dotted short line
        represents the  position of galaxies with  constant mass, with
        the arrow indicating the direction of increasing mass.  }
        \label{fig:kormendy}
   \end{figure}

Figure~\ref{fig:kormendy}   shows  the   relation  between   the  mean
effective  surface brightness  within the  effective radius  ($\langle
\mu_{e,b} \rangle$) against the effective  radius for the S0 bulges in
our  sample.  This  relation is  also known  as the  Kormendy relation
\citep{kormendy77} and  it represents a projection  of the fundamental
plane  \citep{djorgovskidavis87,  dressler87}. \citet{gadotti09}  used
the  Kormendy relation  to  separate classical  from disc-like  bulges
based on the  sensible assumption that they  should be photometrically
and structurally  different.  He suggested that  disc-like bulges must
be faint $\langle \mu_{e,b} \rangle$  outliers of the relation defined
by  ellipticals  and  classical   bulges.   Thus,  he  introduced  the
empirical  line  shown  in  Figure~\ref{fig:kormendy}  as  a  division
between the two  types of bulges.  According only  to this photometric
criterium, and since all our S0  bulges lie in the region of classical
bulges  of the  diagram,  none  of them  would  be  compatible with  a
disc-like  structure.  Nevertheless,  the  Kormendy  relation shows  a
strong     dependence     with     the     spheroid     magnitude/mass
\citep{nigochenetro08} and therefore bulges  below the separation line
might    only   represent    the    less   luminous/massive    systems
\citep{costantin17}.

\subsubsection{Disc properties}
\label{sec:discprop}

Galaxy  discs  in  our  sample  were  fitted  using  either  a  single
exponential profile,  or a double  exponential with a  down-bending or
up-bending outer slope (see  Sect~\ref{sec:photdec}). We found that 17
(68\%), 6 (24\%), and 2 (8\%) S0 galaxies were best fitted with a type
I,  type II,  or type  III  profile, respectively.   These values  are
significantly  different from  those provided  by \citet{erwin08}  for
early type barred galaxies (27\%, 42\%,  and 24\% for types I, II, and
III,  respectively) and  \citet{gutierrez11}  for a  larger sample  of
early  type discs  (30\%, 25\%,  and 45\%  for types  I, II,  and III,
respectively). An obvious source for these differences might be in the
different sample selections and sizes. However, other differences such
as  either the  accurate  selection  of a  well-defined  sample of  S0
galaxies done  in this work or  the application of a  2D decomposition
algorithm to  understand the disc  structure instead of relying  on 1D
azimutally-averaged profiles can also contribute to these differences.
The latter issues are discussed  in detail in \citet{mendezabreu17} and
\citet{ruizlara17}.

\subsubsection{Bar properties}
\label{sec:barprop}

The study of bar  properties is not the main scope  of this paper, but
their  inclusion  in  the   2D  photometric  decomposition  method  is
mandatory to  obtain an accurate  description of the  remaining galaxy
components.   We found  that 21  galaxies  in our  sample are  barred,
representing $\sim  62\%$ of  the sample.  This  value is  higher than
those  found  in  the  literature  for this  range  of  galaxy  masses
\citep{mendezabreu10a,mendezabreu12}  and for  S0  galaxies in  general
\citep{aguerri09,    barazza09},   but    see   \citet{mendezabreu17}.
Recently,  \citet{laurikainen13}  presented  a detailed  inventory  of
photometric structures  in S0 galaxies  finding a strong  variation of
the bar fraction with the bulge prominence.  They found a bar fraction
of $\sim$64\%,  $\sim$64.5\%, and $\sim$32\% for  $B/T$ values between
0-0.2, 0.2-0.4,  and 0.4-1,  respectively.  A  similar trend  has been
recently  reported  by  \citet{buta15}. These  strong  variations  can
explain our high  fraction of bars once the $B/T$  distribution of our
sample is taken into account (22 out of 34 galaxies in our sample have
$B/T < 0.4$).   Moreover, our S0 vs  elliptical separation methodology
is biased  towards barred systems.  Barred  galaxies are automatically
classified as S0 whereas non-barred S0 could still be misclassified as
ellipticals, thus increasing the bar fraction.

Even if not  included in the fit as an  independent component, we also
perform a  visual search for  the presence of 'barlenses'.   A barlens
refers to  the inner part of  a galaxy bar, different  from the bulge,
and they  were first  recognized by  \citet{laurikainen10}.  Recently,
\citet{laurikainen14} and \citet{athanassoula15} use both observations
and numerical simulations to show that  barlenses are likely to be the
more  face-on view  of the  boxy/peanut shape  of the  bar where  seen
edge-on. According to the prescriptions given in those papers we found
signatures of barlenses in 5  barred galaxies in our sample (UGC01271,
NGC1211, NGC1645, NGC3300, and NGC5876)  as well as tentative hints in
other 3 galaxies (NGC0364, NGC4003, and NGC6945).

\subsubsection{Bulge and disc interplay}
\label{sec:bulgedisc}

\begin{figure}
   \centering
   \includegraphics[width=0.49\textwidth]{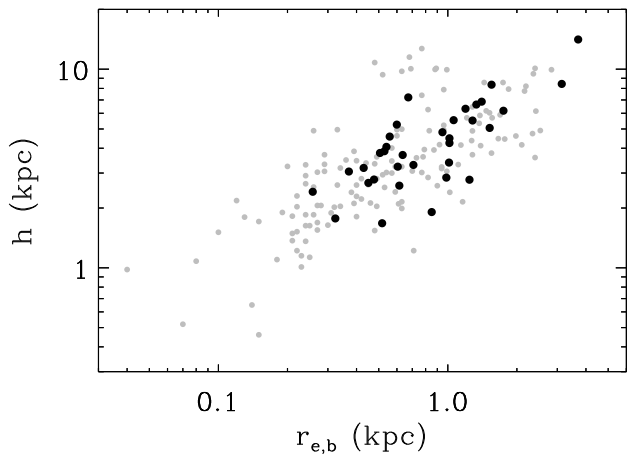}
      \caption{Effective radius of the  bulge vs  scale-length of the
        disc    in   our    S0   sample    (black).    Results    from
        \citet{laurikainen13} are also shown (grey).  }
        \label{fig:revsh}
   \end{figure}

Figure~\ref{fig:revsh} shows the relation between the effective radius
of the bulge and  the scale-length of the disc for  our S0 sample. The
clear  correlation, quantified  using  the  Spearman correlation  test
($\rho \sim 0.7$, statistically significant at $> 3 \sigma$) indicates
that larger bulges reside in galaxies with larger discs. This relation
was already observed by \citet{courteau96}  and later confirmed in the
optical  \citep{aguerri05} and  the near  infrared by  several authors
\citep{mollenhoffheidt01,macarthur03,mendezabreu08a}.     The   values
obtained  from   the  multi-component  photometric   decomposition  of
\citet{laurikainen13} are also shown.   The good agreement between the
different samples  indicate that  despite the small  number statistics
our sample reproduces the expected photometric scaling relation for S0
galaxies.

\subsection{Stellar kinematics of S0 bulges}
\label{sec:rotsupport}
In this section  we analyse the kinematic properties of  our sample of
S0 bulges.  As already stated  throughout the paper, our definition of
bulge  is  entirely  photometric  and  based  on  our  2D  photometric
decomposition.  We use  the value of the  photometric effective radius
to define  the aperture  where the  kinematic parameters  are measured
(see Sect.~\ref{sec:kinprop}), and  we study our S0 bulges  as if they
were an independent structure within  the galaxy.  This assumption has
been widely  adopted in  the literature regarding  either photometric,
kinematic,  or   combined  studies.   An  important   example  is  the
comparison of bulges and other spheroidal systems in scaling relations
related  to  the   virial  theorem  such  as   the  Kormendy  relation
\citep{kormendy85}, Faber-Jackson relation \citep{faberjackson76}, and
the fundamental plane \citep{djorgovskidavis87}.   We refer the reader
to  \citet{falconbarroso16}  for  a  recent review  on  the  kinematic
properties of bulges.

\subsubsection{Disc contamination in our S0 bulges}
\label{sec:disccontamination}

A common caveat associated with the study of the stellar kinematics of
galaxy bulges  is how the  contamination from the underlying  stars in
the disc is affecting the measurements.

\begin{figure}
   \includegraphics[width=0.41\textwidth]{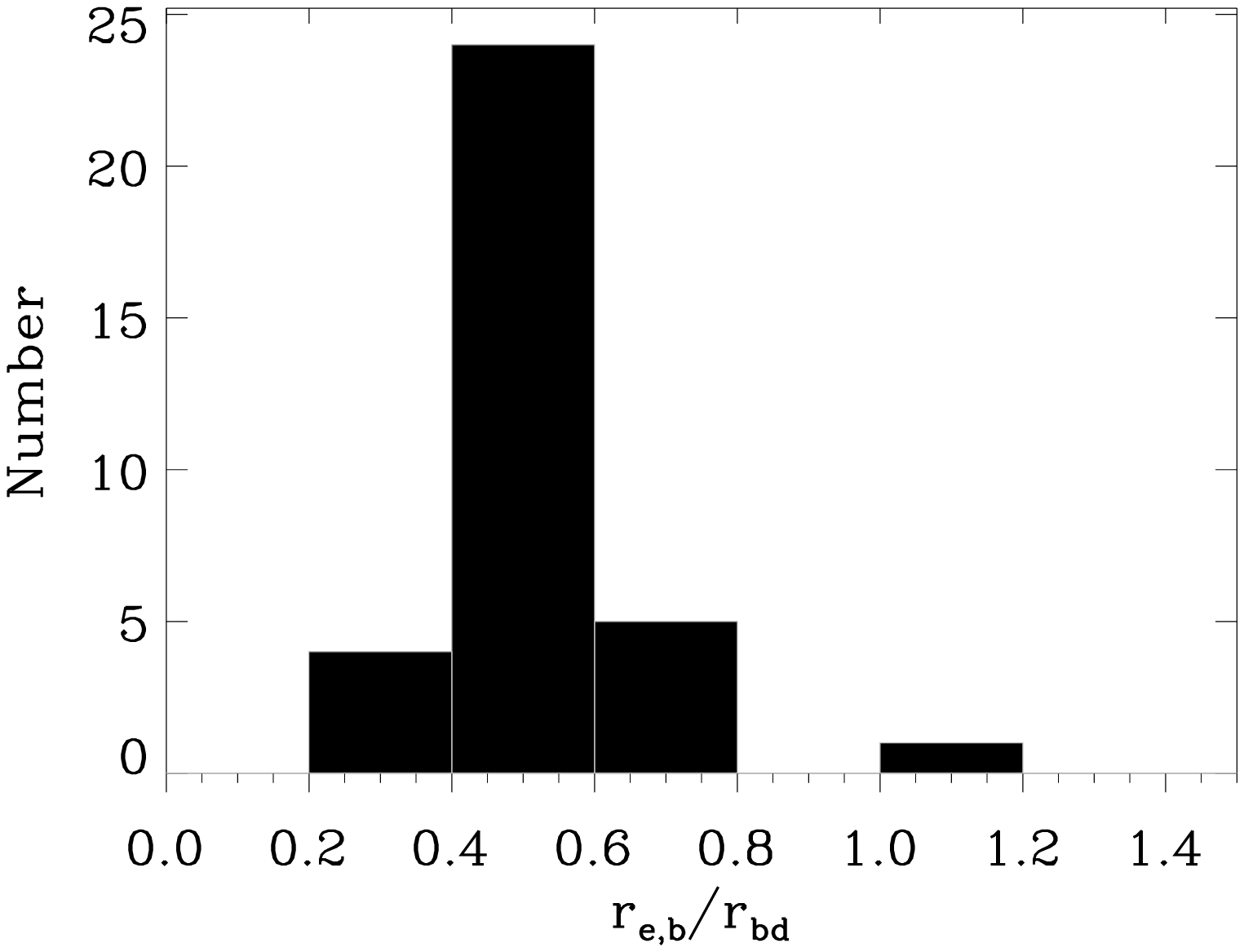}
   \includegraphics[width=0.41\textwidth]{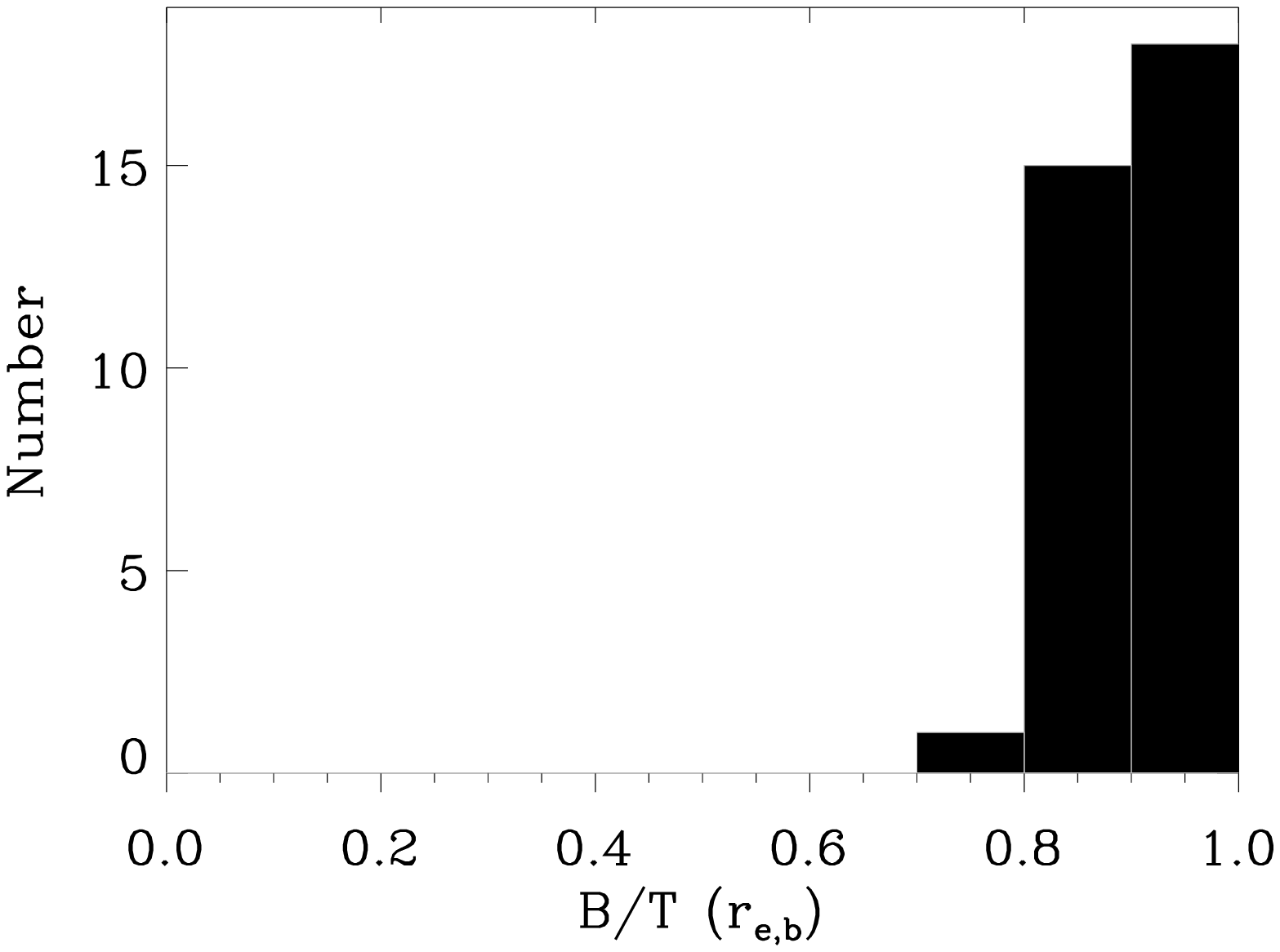}
      \caption{Upper panel.  Distribution  of $r_{e,b}/r_{bd}$ for our
        S0 galaxies.   $r_{bd}$ represents the radius  where the bulge
        light dominates the SB  distribution over any other structural
        component. Bottom panel. Distribution  of $B/T$ ratio computed
        at one $r_{e,b}$ of the bulge.}
        \label{fig:re_rbd}
   \end{figure}

From a  photometric point of  view, we can  quantify the ratio  of the
radial  extension where  the  kinematic  measurements were  performed,
i.e., the  $r_{e,b}$ of the bulges,  with respect to the  radius where
the  light  of another  component  (usually  the  disc) has  the  same
contribution      to     the      SB     distribution      ($r_{bd}$).
Figure~\ref{fig:re_rbd}  (upper  panel)   shows  the  distribution  of
$r_{e,b}/r_{bd}$ values.  It is clear that for most of our galaxies we
are measuring the bulge stellar kinematics within the region dominated
by light  coming from  the bulge.   Similarly, Figure~\ref{fig:re_rbd}
(bottom panel) shows the $B/T$  ratio computed at one bulge $r_{e,b}$.
This allows us to quantify the  integrated amount of light coming from
the  bulge with  respect to  other  structures present  in the  galaxy
centre (i.e., disc  and/or bar).  In all cases, more  than 70\% of the
light in the  region where we are measuring the  stellar kinematics is
coming from the central bulge.

From the  spectroscopic point of  view, quantifying the impact  of the
disc  stellar light  on  our bulge  velocity  and velocity  dispersion
measurements  is  not  straightforward.   We approach  this  issue  by
performing  simulations  on mock  datacubes  in  a similar  manner  as
explained in Appendix \ref{sec:psf}, but including the photometric and
kinematic presence of an underlying disc. A set of 90 tailor-made mock
datacubes are created for each observed galaxy in our sample.  We used
the   measured   values    of   the   bulge   and    disc   SBD   (see
Tables~\ref{tab:decompbulge},   \ref{tab:decompdisc})   to   reproduce
realistic spaxel  to spaxel  intensity variation within  the datacube.
The photometric properties of the datacubes are kept fixed for all the
90 simulated cubes  for each galaxy, allowing us  to produce realistic
$B/T$ ratios in  the region where the stellar  kinematics are measured
(i.e., $r_{e,b}$).  The velocity and velocity dispersion distributions
were   assumed    to   follow    the   analytical    descriptions   by
\citet{salucci07}  and  an   exponential  profile,  respectively  (see
Appendix~\ref{sec:psf} for details on  the actual implementation). The
same  parameterisation was  used for  the bulge  and disc  components.
These  functional forms  involve the  choice of  a maximum  rotational
velocity  ($v_{\rm max}$),  a spatial  scale of  the velocity  profile
($r_{v}$), a maximum central velocity dispersion ($\sigma_{max}$), and
a scale-length of the velocity dispersion distribution ($r_{\sigma}$).
The  analysis  of  the  rotational velocity  and  velocity  dispersion
distributions for our observed galaxies was carried out using the {\it
  kinemetry} code \citep{krajnovic06}. We create mock datacubes within
the limits  of our  observed galaxies  (Appendix~\ref{sec:psf}). Thus,
our  mock   datacubes  were  created  with   the  following  kinematic
characteristics: [$v_{max,b}$,  $v_{max,d}$] =  [50, 150],  [50, 300],
[150, 150], [200, 300], and [300, 300] in km/s, [$r_{v,b}$, $r_{v,d}$]
=  [5, 5],  [10,  10],  and [15,  15]  in arcsecs,  [$\sigma_{max,b}$,
  $\sigma_{max,d}$]  =  [150,  100]  and   [250,  200]  in  km/s,  and
[$r_{\sigma,b}$, $r_{\sigma,d}$] = [10, 10],  [20, 20] and [30, 30] in
arcsecs. The different  combinations of these pairs of  values for the
bulge  and disc  components produce  our  90 mock  datacubes for  each
galaxy. Our  set of kinematic models  cover extreme cases in  terms of
$v_{\rm  max}$, $r_{v}$,  $\sigma_{max}$,  and  $r_{\sigma}$ for  both
components. Nevertheless,  we check that  both the rotation  curve and
velocity  dispersion profiles  obtained from  this analysis  represent
typical observed profiles for real  galaxies. Then, for each galaxy we
create  a  similar  set  of  mock  datacubes  but  removing  the  disc
component. The  differences between these  two sets of  simulations in
terms of the $v/\sigma$ and $\lambda$ values measured within $r_{e,b}$
for  each  galaxy  tell  us  about the  contamination  from  the  disc
component.

   \begin{figure*}
   \includegraphics[bb=64 325 538 720,width=0.49\textwidth]{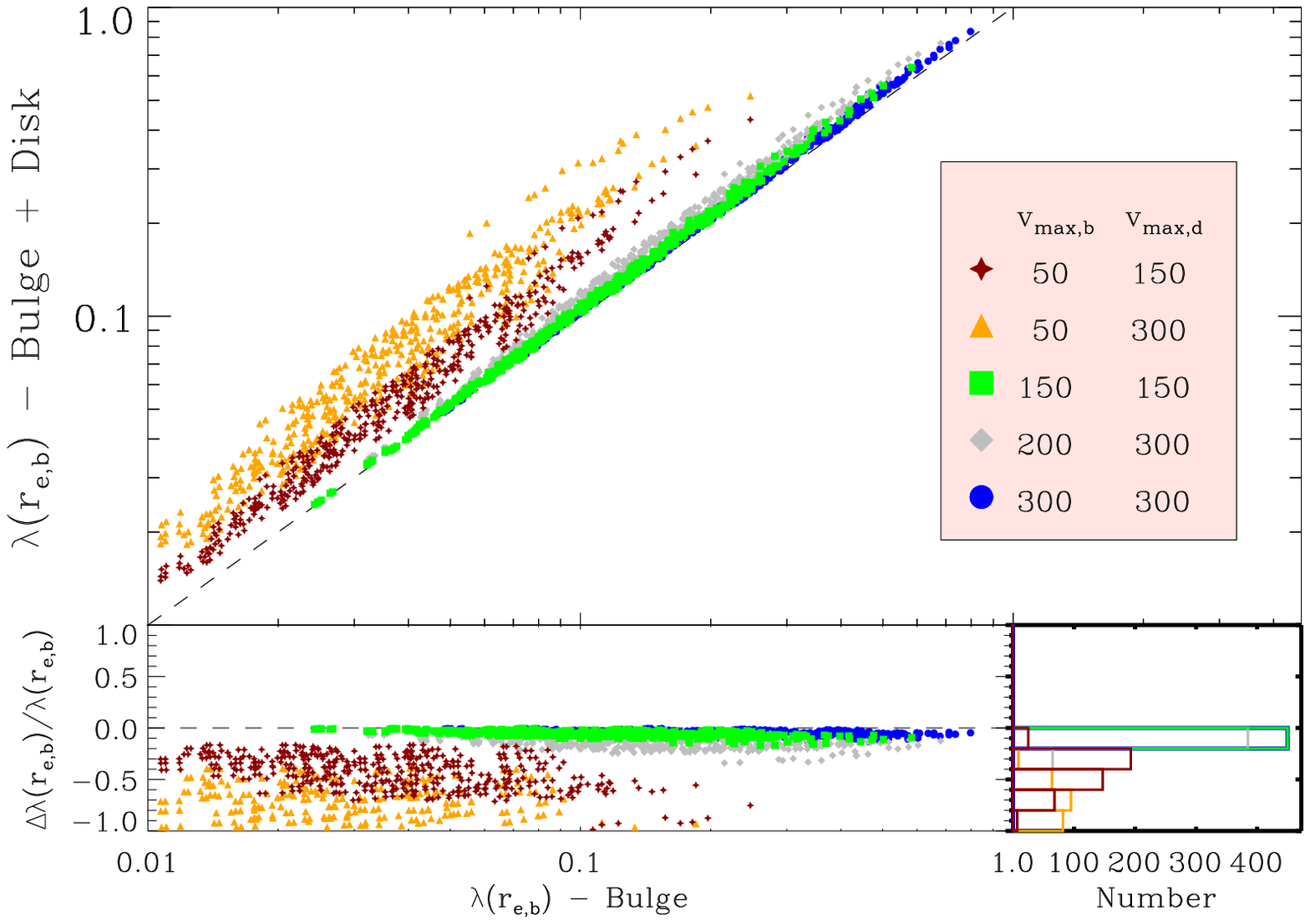}
   \includegraphics[bb=64 325 538 720,width=0.49\textwidth]{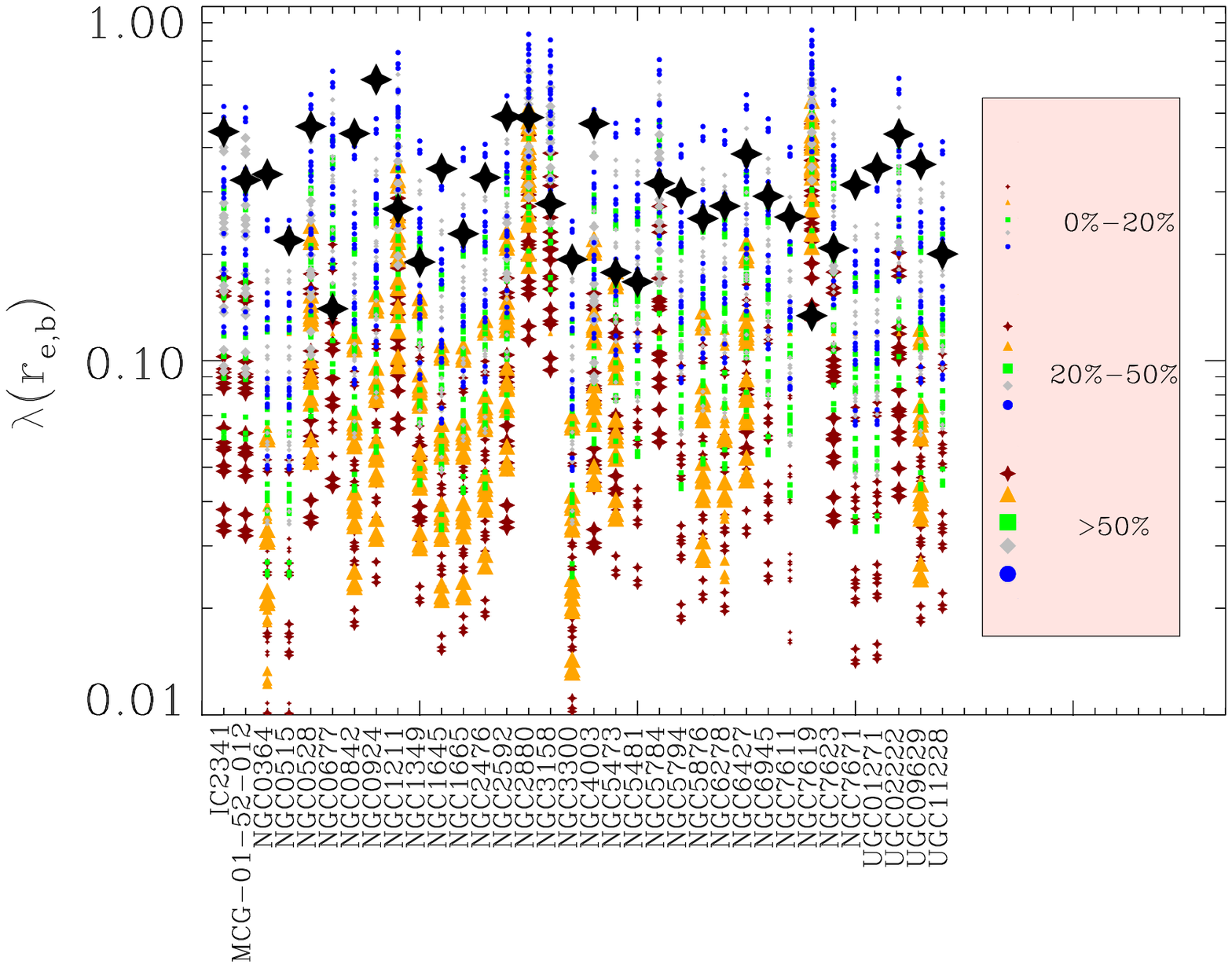}
      \caption{Left panel.  Distribution  of measured $\lambda (r_{\rm
          e,b})$  in  the mock  datacubes  including  a bulge  +  disc
        photometric and kinematic models  against the same measurement
        on mocks including only the bulge model. The different colours
        and symbols  show the  mock datacubes with  the five  pairs of
        maximum rotational velocity (in km/s)  used in this study. The
        lower panel shows the relative  differences for each model and
        the histogram of the differences  for the five different pairs
        of maximum rotational  velocities.  Right panel.  Distribution
        of measured $\lambda  (r_{\rm e,b})$ including a  bulge + disc
        photometric and kinematic models  for each galaxy. Colours and
        symbols as in  the left panel. The size of  the symbol depends
        on the  relative error between the  bulge + disc and  the only
        bulge models. Three different symbol sizes from small to large
        represent relative  differences from 0\%-20\%,  20\%-50\%, and
        $>50\%$, respectively.  The large  black star shows the actual
        measured value  of $\lambda  (r_{\rm e,b})$ for  each observed
        galaxy.  }
        \label{fig:disccontamination}
   \end{figure*}

Figure~\ref{fig:disccontamination}   shows   the    results   of   our
simulations.   The  left  panel  shows the  distribution  of  measured
$\lambda (r_{\rm e,b})$ in the mock datacubes including a bulge + disc
photometric and kinematic model vs the same measurement on simulations
including  only  the bulge  model.   We  separate different  pairs  of
maximum rotational velocities for the  bulge and the disc in different
colours and symbols.  Three different behaviours can be  seen based on
these separations  apart from the  expected larger values  of $\lambda
(r_{\rm  e,b})$   when  the   disc  component   is  included   in  the
modelling. First, datacubes  with low maximum rotational  of the bulge
($v_{\rm max,b}  = $ 50  km/s) show  the largest deviations,  with the
maximum difference depending on the maximum rotational velocity of the
disc. Departures from the actual values of $\lambda (r_{\rm e,b})$ can
be as  high as 100\%,  but most of  the measured values  have $\lambda
(r_{\rm e,b}) <$ 0.1.  Therefore,  bulges with low $v_{\rm max,b}$ are
heavily contaminated by  the underlying disc, but even  in the extreme
case of discs with $v_{\rm max,d} = 300$ km/s they still show $\lambda
(r_{\rm   e,b})$  values   unrealistically  low   compared  with   our
measurements.  The  second trend  is shown  by datacubes  with similar
maximum rotational velocities  for the bulge and the  disc.  They show
almost  no  differences ($<10\%$)  in  the  measured $\lambda  (r_{\rm
  e,b})$, and this  is independent of the  maximum rotational velocity
value.  The third possibility involves  cases where the bulge and disc
maximum rotational  velocity are different,  but the bulge  shows some
rotation ($v_{\rm max,b} = $ 200 km/s).  In this case, the differences
($\sim 20\%$) with  respect to the actual $\lambda  (r_{\rm e,b})$ are
larger than  in the second  case, but much  smaller than in  the first
case.  We conclude that high contamination from the underlying disc is
not strongly dependent on the different maximum rotational velocity of
bulge and disc, but mostly on the rotation of the bulge component.

We  use  these   mock  datacube  simulations  to   quantify  the  disc
contamination in  our measured  $v/\sigma$ and $\lambda$  values.  The
process     is    exemplified     in     the     right    panel     of
Figure~\ref{fig:disccontamination}.   It  shows  the  distribution  of
measured $\lambda (r_{\rm e,b})$, including a bulge + disc photometric
and  kinematic  model,   for  each  galaxy,  with   the  symbol  sizes
representing the deviation  from the input value.   Using those models
with comparable values of $\lambda  (r_{\rm e,b})$ with respect to the
real measurements (i.e, $\left| \delta \lambda (r_{\rm e,b}) \right| <
0.05$), we computed the mean difference and its standard deviation for
each  galaxy.  A similar  approach  was  followed for  the  $v/\sigma$
measurements.  The mean  value of  the difference  is then  used as  a
correction factor for  our measured values of  $\lambda (r_{\rm e,b})$
and $v/\sigma (r_{\rm  e,b})$ and the standard deviation  was added in
quadrature  to the  errors  (see  Sect.~\ref{sec:kinprop}). From  this
analysis,  we  found that  six  of  our  sample bulges  were  strongly
contaminated by  the disc  (large mean value)  and the  correction was
also highly  uncertain (large standard deviation  value), therefore we
decided to remove these bulges from  any further analysis of the bulge
dynamics.  They  are NGC0677, NGC1211, NGC2880,  NGC3158, NGC5784, and
NGC7619.  For the  sake  of completeness  their  kinematic values  are
included in  Table~\ref{tab:kinematics} but not used  in the following
study.

\subsubsection{Schwarzschild dynamical modelling of our galaxy sample}
\label{sec:schwarzschild}

Schwarzschild modelling  of galaxies \citep{schwarzschild79}  has been
demonstrated to be a very powerful  technique to study the dynamics of
stellar   systems   \citep{vandeven06,vandenbosch08}.   Due   to   its
orbit-superposition  methodology,  where  galaxies  are  build  up  by
weighting  the  orbits generated  in  a  gravitational potential,  its
application  to  the modelling  of  real  galaxies  has been  used  to
identify           different            dynamical           components
\citep{vandenbosch08,breddelshelmi14}.

In this paper, we have used  the Schwarzschild modelling of the CALIFA
galaxies carried  out by  \citet{zhu17}.  We refer  the reader  to the
paper  for  a   full  description  of  the  method.    In  short,  the
Schwarzschild  model   requires  an  adequate  model   of  the  galaxy
gravitational  potential   (generally  derived  from   the  luminosity
distribution  of the  galaxy  image). Then,  a  set of  representative
orbits is  explored under  the effect  of this  triaxial gravitational
potential and finally, the combination  of orbits that best reproduces
our galaxy is  found by fitting the observed  luminosity and kinematic
distribution. We  find that 20 out  of 34 galaxies in  our sample were
analysed  using  the Schwarzschild  modelling  by  Zhu et  al.   ({\it
  submitted}).   From these,  we discard  3 of  them because  the disc
contamination is strongly affecting the bulge kinematics, so we remain
with 17 galaxies for this analysis. We use these galaxies to check our
disc contamination correction (Sect.~\ref{sec:disccontamination}), and
to understand our ability to deproject our kinematic measurements.

The Schwarzschild dynamical modelling of  each galaxy provides us with
a  set   of  orbits,   each  one   contributing  differently   to  the
surface-brightness distribution  and stellar kinematics. For  the sake
of comparison with our previous analysis of the real galaxies, we also
looked for  orbits building  both our disc  and bulge  component using
their  luminosity profiles.   We first  determined the  region of  the
galaxy where the disc dominates the SBD of the galaxy according to our
photometric decomposition  (i.e., $r >  r_{bd}$). Then, we  ranked the
orbits by  their relative  contribution to  the disc  total luminosity
(computed  between  $r_{bd}$ and  $r_{max}$,  where  $r_{max}$ is  the
maximum radius used in the Schwarzschild modelling). Finally, the most
luminous orbits contributing up to an  80\% of the total luminosity of
the disc are tagged as belonging to the disc component.  The remaining
orbits were considered to build the bulge component.

\begin{figure}
     \begin{center}
   \includegraphics[width=0.46\textwidth]{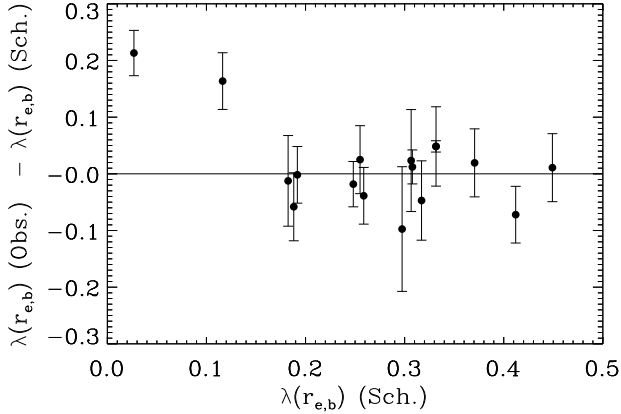}
      \caption{Comparison between the  $\lambda (r_{e,b)}$ values obtained
        from the Schwarzschild dynamical modelling (Sch.) and from our
        measurements using  the disc  correction using  mock datacubes
        (Obs.).  Only the  17  galaxies  with available  Schwarzschild
        modelling are shown.}
        \label{fig:comp_lambda}
     \end{center}
   \end{figure}

Using the previously  defined bulge orbits, we  reconstructed the maps
of SBD,  $v$, and  $\sigma$, and  measured the  $(v/\sigma)_{e,b}$ and
$\lambda_{e,b}$    as    if    they   were    real    galaxies    (see
Sect.~\ref{sec:kinprop}).    Figure~\ref{fig:comp_lambda}  shows   the
comparison  of   $\lambda_{e,b}$  computed  using   the  Schwarzschild
modelling with  respect to our  empirical corrected values  using mock
datacubes.  The agreement between both measurements is remarkable with
most of the  differences being within the estimated  errors (no errors
were  estimated  for  the   Schwarzschild  modelling).   The  standard
deviation  of the  differences is  $\sigma_{Sch.  -Obs.}   \sim 0.08$.
This is  reassuring by  taking into  account the  completely different
methodologies used to remove the disc component.  Still, there are two
galaxies with differences larger than their errors, that correspond to
the lowest values  of $\lambda_{e,b}$ in our sample.   After a careful
check of the orbits derived  from the Schwarzschild modelling, we find
that the  SBD of the disc  orbits do not present  a single exponential
profile, but they are more peaked at the galaxy centre.  The different
slope in the  SBD of the discs  is not taken into account  in our mock
datacubes, and  therefore our  disc correction is  underestimated with
respect to orbital modelling for these galaxies.

\begin{figure}
     \begin{center}
   \includegraphics[width=0.49\textwidth]{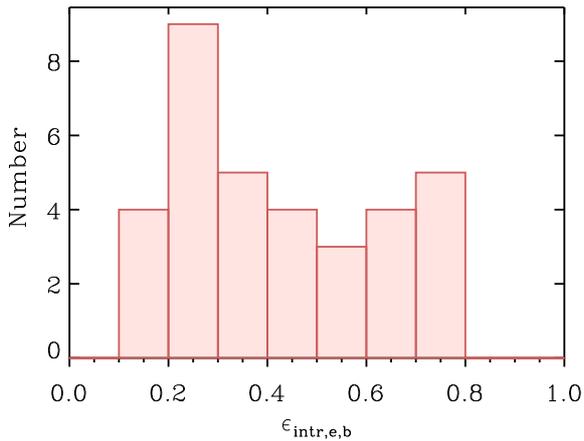}
      \caption{Distribution  of intrinsic  flattening  for our  sample
        bulges. Values  are computed assuming oblateness  for both the
        bulge  and   the  disc.}
        \label{fig:cadep}
     \end{center}
   \end{figure}

\begin{figure}
     \begin{center}
   \includegraphics[width=0.46\textwidth]{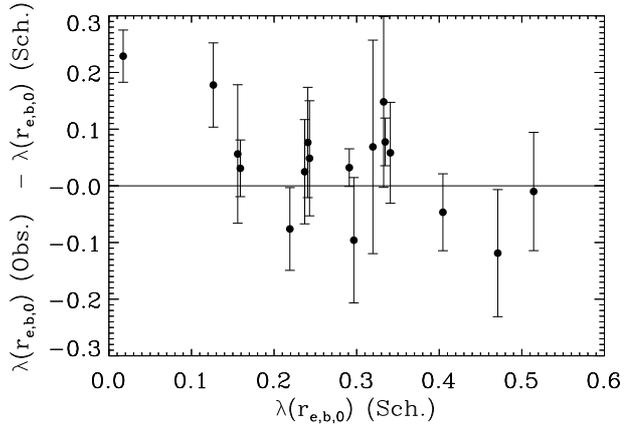}
      \caption{Comparison between the edge-on $\lambda (r_{e,b,0})$ values
        obtained from the Schwarzschild  modelling (Sch.) and from our
        disc correction using mock datacubes (Obs.). The Schwarzschild
        values are directly measured on  the edge-on view of the bulge
        model. The  observed values  are deprojected in  a statistical
        way (see text for details).}
        \label{fig:comp_lambdaedgeon}
     \end{center}
   \end{figure}

Another advantage of  the Schwarzschild modelling is that  we have now
the  possibility of  measuring  the values  of $(v/\sigma)_{e,b}$  and
$\lambda_{e,b}$   using   the   edge-on  projection   of   the   bulge
model. Observationally, the measured  values of $(v/\sigma)_{e,b}$ and
$\lambda_{e,b}$ depend on  three parameters of the  bulge: the orbital
anisotropy, the intrinsic  shape, and the inclination  with respect to
the  line-of-sight  \citep[see][]{emsellem11}.   Observations  do  not
provide  access to  the orbital  anisotropy. Therefore,  we considered
that the vertical anisotropy of our  bulges can take any value from $0
<  \beta <  1$,  and we  added  this uncertainty  to  the error  bars.
Regarding the intrinsic flattening, we  considered that both the bulge
and disc are oblate ellipsoids  sharing the same inclination.  Despite
this  being  a   strong  assumption  \citep[see][]{mendezabreu16},  it
provides  a first  order estimation  that helped  us to  deproject the
bulge  kinematics.  The  distribution of  intrinsic flattening  of our
bulges is  shown in  Figure~\ref{fig:cadep}.  Finally, we  derived the
galaxy inclination  assuming that  discs have an  intrinsic flattening
given  by a  normal  distribution with  mean  flattening $\langle  C/A
\rangle$   =  0.267   and   standard  deviation   $\sigma_{C/A}$ =0.102
\citep[see][]{rodriguezpadilla13}.  Figure~\ref{fig:comp_lambdaedgeon}
shows the  comparison of the edge-on  $\lambda_{e,b,0}$ values derived
from  the Schwarzschild  modelling  and  our deprojected  measurements
based on  the observations.  Despite  the larger errors,  the standard
deviation  of the  differences is  $\sigma_{Sch.0 -Obs.0}  \sim 0.09$,
showing a good  agreement. Similarly as for the  projected values, the
two  galaxies  with  lower  $\lambda_{e,b,0}$ also  show  the  largest
differences.

A similar  analysis was  performed for $(v/\sigma)_{e,b}$  obtaining a
standard deviation of the differences $\sigma_{Sch.  -Obs.}  \sim 0.1$
and  $\sigma_{Sch.0   -Obs.0}  \sim   0.12$  for  the   projected  and
deprojected  values, respectively.   The  good  agreement between  the
results from our empirical  correction and the Schwarzschild modelling
confirms our ability  to recover the bulge stellar  kinematics. In the
following, we will consider only  the deprojected values obtained from
our  statistical  analysis unless  otherwise  stated.   Using our  own
estimation allows  us to use the  whole sample of 28  bulges with good
kinematics  described throughout  the  paper.   The final  deprojected
values are listed in Table~\ref{tab:kinematics}.

\subsection{Photometry vs kinematics in S0 bulges}
\label{sec:photkin}

\begin{figure*}
     \begin{center}
   \includegraphics[bb=54 340 558 720,width=0.69\textwidth]{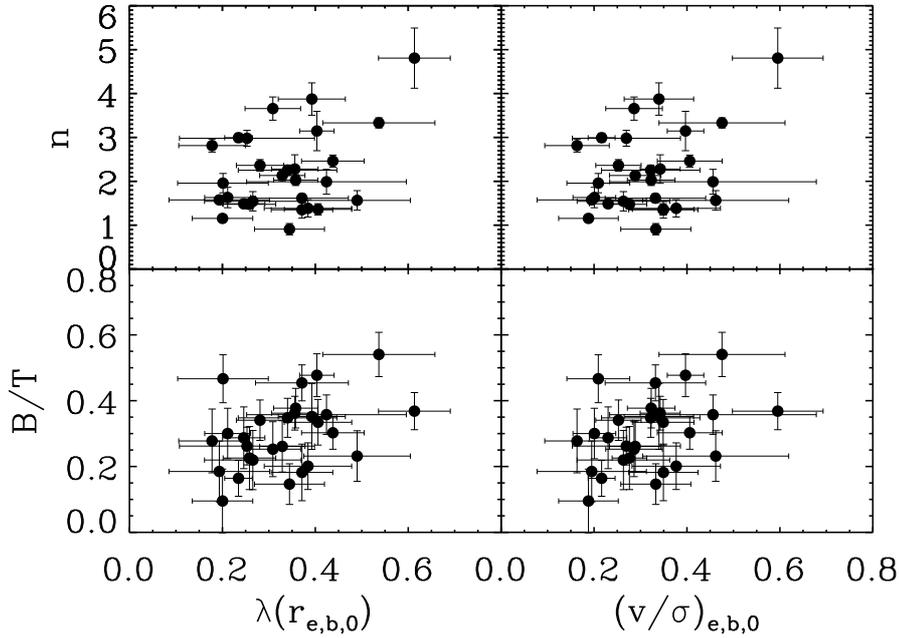}
      \caption{{\it Upper panels}. The left  and right panels show the
        distribution of the  bulge S\'ersic index $n$  with respect to
        the   deprojected   values   of  $\lambda   (r_{e,b,0})$   and
        $(v/\sigma)_{e,b,0}$, respectively. {\it  Bottom panels.}  The
        left   and  right   panels  show   the  distribution   of  the
        bulge-to-total luminosity  ratio ($B/T$)  with respect  to the
        deprojected    values    of    $\lambda    (r_{e,b,0})$    and
        $(v/\sigma)_{e,b,0}$, respectively.}
        \label{fig:morphokin}
     \end{center}
   \end{figure*}

Figure~\ref{fig:morphokin}  shows  the   distribution  of  deprojected
values of $\lambda_{e,b,0}$ and $(v/\sigma)_{e,b,0}$ for our S0 bulges
sorted by their photometric  properties: S\'ersic index (upper panels)
and $B/T$ (bottom panels). We do  not find any clear trend between the
photometric and  kinematic properties  of our  bulges. We  compute the
Spearman  correlation  test  ($\rho$)   in  order  to  understand  the
statistical  significance  of  a   possible  correlation  between  the
proposed measurements.   This test assesses how  well the relationship
between  two  variables  can  be  described  using  a  monotonic  (not
necessary  linear)  function.   We  find  low  values  of  the  $\rho$
correlation coefficient in  all cases: 0.2, 0.4, 0.2, and  0.4 for the
$\lambda_{e,b,0}$    vs.    $n$,    $\lambda_{e,b,0}$   vs.     $B/T$,
$(v/\sigma)_{e,b,0}$  vs.  $n$,  and  $(v/\sigma)_{e,b,0}$ vs.   $B/T$
relations,  respectively.  In  addition, we  computed the  statistical
significance  of the  possible correlation  with respect  to the  null
hypothesis (no  correlation). We find  that we cannot reject  the null
hypothesis at more  than 1 $\sigma$ in any of  the cases, therefore we
can  conclude that  there is  no statistical  correlation between  the
photometric  ($n$  and  $B/T$) and  kinematic  ($\lambda_{e,b,0}$  and
$(v/\sigma)_{e,b,0}$) properties of our bulges.

Due   to  the   relatively  large   uncertainties  of   the  kinematic
measurements for some  of our bulges, we decided to  perform a further
test   to    understand   the   statistical   significance    of   our
results. Therefore, we carried out  the Spearman test using Monte Carlo
simulations taking into account the  errors in both variables for each
case  (i.e.,  $\lambda_{e,b,0}$  or $(v/\sigma)_{e,b,0}$  and  $n$  or
$B/T$).  We performed  1000 simulations allowing each bulge  to take a
possible value  confined within its  error, and computed  the Spearman
correlation coefficient for each simulation of the sample. As a result
of this  exercise we obtained  a distribution of both  the correlation
coefficients ($\rho$) and statistical significance. We found that, for
correlations including the  S\'ersic index, only in 3\%  of the trials
the  null hypothesis  could be  rejected  at 2  $\sigma$. This  number
increase up  to a 10\%  regarding the correlations with  $B/T$.  These
percentages  are low  and they  confirm that  there is  no correlation
between the photometry  and kinematics of our  bulges independently of
the uncertainties in the measurements.

In Sects.  \ref{sec:disccontamination}  and \ref{sec:schwarzschild} we
discussed  how  the disc  contamination  was  removed from  the  bulge
kinematics.  We make  use of the kinematics results  obtained from the
Schwarzschild  reconstructed   bulge  maps   to  study   the  possible
correlation  with their  photometric  properties.   Despite the  lower
number statistics (17 galaxies), we  confirm results obtained with the
whole sample about the lack of a statistically significant correlation
between the morpho-kinematic properties of S0 bulges.

\begin{figure*}
     \begin{center}
   \includegraphics[bb = 54 400 558 610, width=0.8\textwidth]{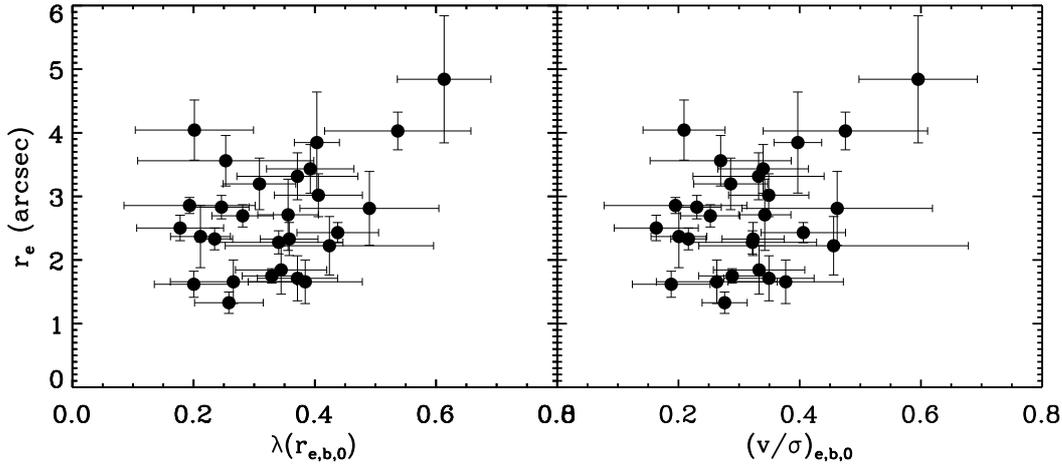}
      \caption{The left and  right panel show the  distribution of the
        deprojected      values      of     $\lambda_{e,b,0}$      and
        $(v/\sigma)_{e,b,0}$  with  respect   to  the  apparent  bulge
        effective radius ($r_e$), respectively.}
        \label{fig:kin_re}
     \end{center}
   \end{figure*}

Figure~\ref{fig:kin_re} shows the relation  between the apparent bulge
effective   radii    ($r_e$)   and    the   deprojected    values   of
$\lambda_{e,b,0}$  and  $(v/\sigma)_{e,b,0}$  for our  S0  bulges.   A
possible caveat  to our analysis might  be the small apparent  size of
our  bulges.   We  checked  whether  the  kinematic  measurements  are
correlated with  the aperture  where they  were measured  ($r_e$).  We
carried  out  the  same  statistical  analysis,  based  on  Monte Carlo
simulations, as performed  with the $n$ and $B/T$  parameters. We find
that we can only reject the lack of correlation at 2 $\sigma$ level in
less than  10\% of the  realisations, showing the lack  of correlation
between our kinematic measurements and bulge apparent size.

The  relation  between the  kinematic  and  photometric properties  of
bulges has  been greatly debated in  the literature, but it  is poorly
constrained by observations.  Previous observational results suggested
that bulges with low S\'ersic index  and low $B/T$ should present more
disc-like characteristics \citep{fisherdrory16}, therefore they should
show larger rotational support than bulges with either higher S\'ersic
index or $B/T$.   We have demonstrated that there is  no such trend in
our sample of S0 bulges.

\begin{figure}
     \begin{center}
   \includegraphics[bb=54 380 558 720,width=0.49\textwidth]{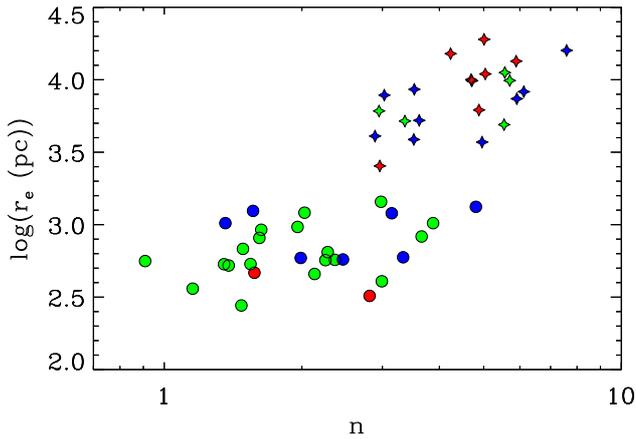}
      \caption{Distribution of the bulge  effective radius ($r_e$) vs.
        S\'ersic    index     ($n$)    for    our     sample    bulges
        (circles). Elliptical galaxies are shown with stars. The colour
        code  represent bulges/ellipticals  with  different values  of
        $\lambda_{e,b,0}$:  blue,  green,  and  red  colours  describe
        bulges/ellipticals   with  $\lambda_{e,b,0}   <  0.2$,   $0.2<
        \lambda_{e,b,0}   <  0.4$,   and   $\lambda_{e,b,0}  >   0.4$,
        respectively. }
        \label{fig:n_re_lambda}
     \end{center}
   \end{figure}

Another  commonly  used  diagram  to separate  bulges  with  disc-like
properties from bulges with features similar to ellipticals (classical
bulges)  is presented  in  Figure~\ref{fig:n_re_lambda}.   It shows  the
relation between  the S\'ersic index  and the effective radius  of the
bulge.   In general,  bulges with  both low  $n$ and  low $r_{e}$  are
considered disc-like whereas  bulges with large $n$  and large $r_{e}$
are considered  as classical  \citep{fisherdrory10}.  In order  to add
information about their kinematic properties  we have included a colour
code  where  blue,  green,  and  red  circles  represent  bulges  with
$\lambda_{e,b,0}   <  0.2$,   $0.2<   \lambda_{e,b,0}   <  0.4$,   and
$\lambda_{e,b,0} > 0.4$, respectively. We also included a subsample of
elliptical  galaxies  (see Sect.~\ref{sec:globalprop})  with  measured
kinematics at 1 $r_e$ of the galaxy.  We do not find any trend between
the stellar kinematics and the position of the bulges in this diagram.
Previous works reported the presence of a break (knee) separating both
kind  of bulges  \citep{fisherdrory16}, we  find that  for our  sample
bulges there  is no such break,  but this is only  present whenever we
add the elliptical galaxies to  the sample. 

Therefore,  we  suggest that  pure  photometric  diagrams, such  as  a
S\'ersic index  based separation or  the $r_e$ vs $n$  relation, might
not  be useful  to  separate  disc-like from  classical  bulges in  S0
galaxies.

\section{Discussion}
\label{sec:discussion}

\subsection{Morpho-kinematic properties of S0 bulges}

The  photometric  properties  of  our  S0  bulges  (effective  radius,
S\'ersic index,  and $B/T$  ratio) span  a wide  range of  values (see
Figure   \ref{fig:bulphot}).   This   lack  of   homogeneity  in   the
photometric properties of S0 bulges has raised the idea that different
formation  scenarios, or  at least  different initial  conditions, are
needed to explain their observational properties.

From the  kinematics point  of view,  we find that  our bulges  show a
large  range   of  angular  momentum  and   ($v/\sigma$)  values.   We
demonstrate  that  this  result  is robust  despite  the  difficulties
inherent to separate  the bulge kinematics from the  total galaxy, and
the uncertainties inherent to deprojection issues.

Figure~\ref{fig:morphokin}  shows   our  attempt  to   understand  the
possible  connection  between  bulge photometry  and  kinematics.   We
demonstrate  that  regardless  of  projection  effects,  there  is  no
statistically significant relation between  either the S\'ersic index,
or the  $B/T$ luminosity ratio,  and the stellar angular  momentum, or
the $v/\sigma$, of  our S0 bulges.  A similar lack  of correlation was
found  in  \citet{falconbarroso03}.   They  showed  that  the  central
gradient of  the velocity dispersion  was not related to  the S\'ersic
index for a sample of 19 early-type disc galaxies.  On the other hand,
\citet{fabricius12} claimed that purely kinematic diagnostics of bulge
dichotomy agree with those based on S\'ersic index, i.e., low S\'ersic
index bulges have increased rotational support.  This morpho-kinematic
approach has barely  been explored in the literature,  and the results
of \citet{fabricius12}  show a  significant degree of  overlap between
different  bulge  types.  Nevertheless,  we  suggest  that a  possible
explanation of  the results is  the different morphological  mixing in
\citet{fabricius12}  sample  ($\sim70\%$  spiral,  $\sim$30$\%$  S0s),
suggesting a  diverse origin of  their bulges, possibly with  a larger
fraction of  disc-like bulges in  later Hubble types, with  respect to
our  pure  sample  of  S0s. Recently,  \citet{tabor17}  performed  the
spectro-photometric bulge to disc decomposition of three galaxies from
the   CALIFA  survey.    They  found   that  all   their  bulges   are
photometrically  described by  a  S\'ersic  index $n\sim1$.   However,
despite the stellar kinematics of  the bulges show some rotation, they
are  considered as  dispersion  dominated systems  due  to their  high
velocity dispersion values.

\subsection{Bulge formation scenarios}
\label{sec:bulgeformation}
Different  formation   scenarios  are   expected  to   leave  distinct
fingerprints  on the  photometric and  kinematic properties  of bulges
\citep{kormendykennicutt04,athanassoula05}.    These   scenarios   are
generally divided into two main types depending on whether their final
outcome could be classified as a classical or disc-like bulge. 

The  major merger  scenario  has  commonly been  invoked  as the  main
channel for  the formation  of classical-like bulges  \citep[][but see
  \citealt{keselmannusser12}]{hopkins09}.    
Recent merger simulations in  a cosmological context have demonstrated
that  only the  right  combination of  major and  minor,  wet and  dry
mergers   can   reproduce   the   current   population   of   galaxies
\citep{oser12}.  In  particular, the  amount of  gas available  in the
merger   has   been    found   to   be   critical    for   the   bulge
growth. \citet{hopkins09}  showed that  dissipative processes  are the
main  driver  of   bulge  growth  after  a   merger,  whereas  stellar
redistribution plays a minor  role \citep{brooks16}.  These mechanisms
have been  explored in high resolution  cosmological simulations where
the   properties   of   the   bulge  regions   are   better   resolved
\citep{guedes13,okamoto13,  christensen14}.  If  bulges are  formed by
mergers, like elliptical galaxies, then  they should show a variety of
photometric and kinematic properties depending mainly on the amount of
dissipation  involved  in  the merger  and,  eventually,  merger-built
bulges   might  show   the   decoupled   morpho-kinematics  found   in
Figure~\ref{fig:morphokin} \citep{hopkins10,naab14}.

At  high  redshift,  disc-like star-forming  galaxies  have  irregular
optical  morphologies  dominated by  giant  clumps  of star  formation
\citep{abraham96,  vandenbergh96,elmegreen07,hinojosagoni16}.  These
clumpy  galaxies might  be the  early progenitors  of the  S0 galaxies
observed in the nearby Universe.   
The observed properties of the  bulges resulting from this process are
still a  matter of debate.   However, some numerical  simulations have
already  provided some  ideas about  the  final outcome  of the  clump
merging  process.   If the  clumps  are  short-lived, and  efficiently
destroyed by stellar feedback before their inward migration timescale,
there could still be a diffuse inflow of inter-clump gas driven by the
instability   \citep{hopkins12,bournaud11}.    This  will   create   a
low-concentration  bulge if  no other  relaxation process  affects the
central region.  On the other hand, the models with long-lived clumps,
the repeated clump coalescence \citep{elmegreen08}, and the relatively
short star-formation timescales  \citep{immeli04} would produce bulges
with low rotational  support and high S\'ersic indices.   While we are
far from  a definitive  answer about  the observational  properties of
clump-driven bulges and their evolution until $z\sim0$, the variety of
morphological and  kinematic properties  predicted from  recent models
might also be in agreement with the results shown in this paper.

The minor merger (mass ratios lower  than 1:4) mechanism is also known
to  induce gentle  transformations to  the prominence  of the  remnant
bulge \citep{aguerri01,elichemoral06,elichemoral11},  producing in all
cases an increase of the S\'ersic index. The stellar kinematics of the
bulge remnants after  dry minor merger accretion  of galaxy satellites
was studied  by \citet{tapia14}.   They found that  the net  effect of
multiple minor mergers  is to increase (or keep)  the $v/\sigma$ ratio
of  the  bulges.  Therefore,  the combination  of  the  kinematic  and
photometric  evolution of  bulges  due  to minor  mergers  leads to  a
scenario  where there  might  be  a lack  of  correlation between  the
morpho-kinematics of  the remnant  bulges, and in  extreme cases  to a
scenario where  high S\'ersic  index bulges  might have  large angular
momentum values.

Internal secular  evolution processes,  such as  those induced  by the
presence of a bar  or spiral arms in an unstable  disc, are thought to
produce   disc-like  bulges   \citep{kormendykennicutt04}.   In   this
picture, non-axisymmetric galaxy structures  such as bars redistribute
the angular momentum of disc material and thereby they are responsible
for  an efficient  transport  of  gas into  the  central bulge  region
\citep{pfenningernorman90,friedlibenz95}.   The stellar  kinematics of
these disc-like bulges is expected to  be dominated by rotation with a
structure similar to an oblate ellipsoid.  Most of the galaxies in our
sample host a  stellar bar (72\%) and we find  some low-S\'ersic index
bulges  with   relatively  large   values  of   $\lambda_{e,b,0}$  and
$(v/\sigma)_{e,b,0}$ in our sample that might be considered compatible
with an internal secular evolution origin.  We investigated a possible
relation  between the  presence of  a bar  and the  rotational support
($\lambda_{e,b,0}$ and  $(v/\sigma)_{e,b,0}$) of  our bulges  but none
was found.
The  lack  of  relation  could be  explained  if  bar-induced  secular
evolution is  not efficient in  building new central structures  in S0
galaxies. This  scenario needs either  a deficit  of gas in  the outer
disc  \citep[see][]{masters10}  or  an   inefficient  gas  inflow  and
posterior   star    formation.    \citet{delorenzocaceres13}   already
suggested  a  minor  role  of bar-induced  secular  evolution  in  the
formation of new  central structures.  In addition,  recent results by
\citet{seidel15} confirmed that  more than 50\% of the  bulge mass was
created at high redshift and not due to secular evolution.

Based on the morpho-kinematic properties  derived for our sample of S0
bulges it seems unlikely that  they were mainly formed through secular
processes induced by bars.  On the other hand, dissipational processes
taking  place  at  high  redshift  such as  major  galaxy  mergers  or
coalescence  of  star-forming  clumps   are  favoured.   The  relative
influence of these two mechanisms and the role of minor mergers on the
observed  properties  of our  bulges  is  difficult to  evaluate,  and
further  high-resolution cosmological  simulations  are necessary  to
address this problem.

\subsection{Implications on S0 formation}
\label{sec:s0formation}

In  Sect.~\ref{sec:globalprop}  we  discussed  the  local  and  global
environment where  the S0 galaxies in  our sample live. We  found that
none of  the S0s  belong to  a known cluster  structure, and  that the
majority of  our S0 galaxies  live in a low-density  local environment
similar to  that of the  field or loose  groups. Most of  the proposed
mechanisms  able to  transform  a star-forming  spiral  galaxy into  a
passive S0 are related to high-density environments: galaxy harassment
\citep{moore96,moore99,aguerrigonzalezgarcia09},      ram     pressure
stripping        \citep{gunngott72,quilis00,bekki09},       starvation
\citep{bekki02}, or  gravitational heating \citep{khochfarostriker08}.
Therefore, a  suggested path of  S0 formation invoking a  quenching of
star-formation through any of these  mechanisms, and a later fading of
spiral galaxies into S0s is unlikely  to be happening in our galaxies.
In the fading  scenario due to high-density environments,  most of the
previous  processes only  affect either  the gas-phase  of the  galaxy
\citep[i.e.,  ram pressure  stripping, starvation;][]{abadi99}  or the
outer              stellar             discs              \citep[i.e.,
  harassment][]{aguerrigonzalezgarcia09}, therefore  leaving unchanged
the  stellar  angular momentum  of  the  galaxy. Bulges  of  late-type
spirals  have  been  found  to  be  well  represented  by  exponential
surface-brightness \citep{laurikainen10}  and with  stellar kinematics
typical of a disc-like  structure \citep{ganda06}.   As previously
  shown, we found that the morphology  and kinematics of our bulges are
  not correlated, with photometrically  exponential bulges not showing
  the largest rotational support, pointing against a simple transition
  from spirals to  S0 in our sample.  A similar  argument in terms of
the   global  galaxy   properties  has   recently  been   proposed  by
\citet{querejeta15b}.  They  claim that  the stellar  angular momentum
and concentration  of late-type spiral galaxies  are incompatible with
those of S0s, thus concluding  that fading alone cannot satisfactorily
explain the evolution from spirals into S0s.

The  star-formation   quenching  of  spiral  galaxies   is  a  process
associated not  only to high-density  environments, but could  also be
related  to  galaxy   internal  processes  \citep[see][and  references
  therein]{aguerri12}.   AGN   feedback  has  been  suggested   to  be
efficient in  transforming early-type  spiral galaxies located  in the
blue  cloud into  galaxies located  close to  or on  the red  sequence
\citep{schawinski06}.  However,  stellar angular momentum  studies are
usually  not  compatible  with   a  simple  quenching  mechanism,  and
dynamical  evolution   of  the  system   is  needed  to   explain  the
observations \citep{querejeta15b}. Another  internal mechanism able to
modify the star formation and dynamics of the galaxy centre is related
to bar structures.  If gas is efficiently driven by bars to the galaxy
centre, it could  accelerate the depletion of the gas  supply from the
outer disc. If this process is  not balanced by an increased inflow of
cosmological gas, this would ultimately produce a quiescent red barred
galaxy  as those  studied  in  this paper  \citep{cheung13,masters11}.
Nevertheless the  likely massive disc-like bulges  resulting from this
process \citep{kormendykennicutt04} are not the main population in our
sample.  Still,  unresolved inner  rotating structures could  still be
present in our galaxies and this scenario cannot be ruled out.

We have discussed  the most likely formation scenarios  for our sample
of field S0  galaxies.  {\it We suggest that the  global properties of
  our S0  galaxies also  point toward a  formation mechanism  based on
  dissipational processes  at high  redshift; either major  mergers or
  gas accretion  onto gravitationally  unstable disc galaxies,  with a
  possible  later  evolution due  to  minor  merger accretion.}   This
picture  is  also consistent  with  a  mass-related evolution  of  S0s
\citep{vandenbergh09,barway13}  and   the  general  picture   for  the
evolution  of early-type  galaxies  proposed in  \citet{cappellari13}.
According to this  view, massive S0 galaxies have likely  formed at an
early epoch  through major mergers, as  it is believed to  be the case
with elliptical galaxies.   On the other hand, faint  S0 galaxies have
likely formed through secular processes.  We cannot discard that faint
S0s  would have  been originated  from  spiral galaxies  which in  the
process of  their interaction with  dense environments had  their star
formation      quenched     due      to      stripping     of      gas
\citep{aragonsalamanca06,barr07}.   Our sample  of S0  galaxies target
only the high  mass end of the S0  family (see Fig.~\ref{fig:galprop})
and therefore would be compatible  with this mass-dependent idea of S0
formation.  

\section{Conclusions}
\label{sec:conclusions}

We have studied  the photometric and kinematic properties  of a sample
of 34 S0 bulges drawn from  the IFS CALIFA survey.  Extensive work has
been  devoted  to  select  our  final sample  of  {\it  bona-fide}  S0
galaxies.    We   developed  a   two-step   method   to  identify   S0
galaxies. First, all visually classified elliptical and S0 galaxies in
the CALIFA sample  pass through a {\it logical filtering}  in order to
provide the best  fit model with a physical meaning.   Then, the final
model selection  was done using  the $BIC$ statistical  criteria.  The
aim  was  to  obtain  a  well-defined sample  of  {\it  canonical}  S0
galaxies,  i.e.,  composed  by  a  central bulge  and  an  outer  disc
dominating the light in the galaxy outskirts.

Our final sample  was found to be representative of  a particular type
of S0s. All  galaxies have high stellar  masses ($M_{\star}/M_{\sun} >
10^{10}$), they lie  on the red sequence, and they  live in relatively
isolated  environments with  a local  density similar  to that  of the
field and loose groups.

A careful multi-component photometric  decomposition of the sample was
performed to  derive the bulge  parameters using the GASP2D  code. The
structural parameters of  the S0 bulges were used to  both provide the
galaxy region  from which the  stellar kinematics were  extracted and,
combined  with  the  galaxy   dynamics,  to  constrain  the  formation
scenarios of  S0 bulges. 

Using the CALIFA  IFS data we have explored the  stellar kinematics of
our S0 bulges measuring the  $v/\sigma$ vs $\epsilon$ and $\lambda$ vs
$\epsilon$ diagrams  within 1 $r_{e,b}$  of the bulge.   We quantified
the impact of  the underlying large scale disc in  the bulge kinematic
measurements using both mock spectroscopic datacubes and Schwarzschild
dynamical modelling of the galaxies.  We  found that six bulges in the
sample  were  heavily contaminated  and  they  were removed  from  the
analysis.  The remaining bulges (28 galaxies) were corrected from disc
contamination and deprojected  to the edge-on line of  sight using the
statistical approach.

We found  that our  S0 bulges  show a  large range  of values  for the
deprojected $\lambda_{e,b,0}$ and $(v/\sigma)_{e,b,0}$ values.
We also found  a lack of correlation between the  photometric ($n$ and
$B/T$) and kinematic  ($v/\sigma$ and $\lambda$) properties  of the S0
bulges. This  behaviour might  be puzzling in  the current  picture of
bulge   formation  where   classical   bulges  are   expected  to   be
photometrically  modelled with  high  S\'ersic  index $n  >  2$ and  a
kinematics dominated by random motions, whereas secularly-built bulges
are  expected  to  be  disc-like  in  both  their  light  distribution
($n\sim1$) and kinematics (dominated by  rotation). We have found that
purely  photometric diagnostics  separating  disc-like from  classical
bulges  might not  be applicable  to S0  bulges. We  discuss that  the
observed photometric and stellar  kinematic properties of the majority
of  our S0  bulges  are  hard to  reconcile  with  the predictions  of
numerical  simulations  for  an internal  secular  evolution  scenario
driven by bars or spiral arms.

In summary, we suggest that  the morpho-kinematic properties of our S0
bulges  might be  explained if  field S0  galaxies were  mainly formed
through dissipational processes  happening at an early  stage of their
evolution, either  through wet major  mergers or coalescence  of giant
star-forming  clumps.  Then,  a possible  later evolution  might imply
that  galaxies evolved  secularly through  both external  accretion of
satellite  galaxies (inducing  changes  in the  bulge properties)  and
internal  bar-induced mechanisms  in gas-devoided  discs (with  little
effect in the formation of new central structures). These results seem
to also  be supported  by the  global properties  of our  S0 galaxies,
i.e., their high masses and relatively isolated environment.  However,
a  proper  understanding  of  the dominant  formation  process  of  S0
galaxies  will  not  be achieved  until  high-resolution  cosmological
simulations  resolving  the bulge  structure  and  kinematics will  be
studied.

\section*{Acknowledgements}

We  would  like   to  thank  the  referee  for   useful  comments  and
suggestions.   JMA  and  VW  acknowledges support  from  the  European
Research Council  Starting Grant (SEDmorph;  P.I.  V.  Wild).  JMA and
JALA acknowledge  support from  the Spanish  Ministerio de  Economia y
Competitividad (MINECO) by the  grant AYA2013-43188-P.  J.  F-B.  from
grant AYA2016-77237-C3-1-P  from the  Spanish Ministry of  Economy and
Competitiveness  (MINECO).  J.   F-B  and  GvdV acknowledge  finantial
support from the  FP7 Marie Curie Actions of  the European Commission,
via  the Initial  Training  Network DAGAL  under  REA grant  agreement
number 289313.  TRL  and EF acknowledge the support  from the projects
AYA2014-53506-P and JA-FQM-108. AdLC  acknowledges support from the UK
Science and  Technology Facilities  Council (STFC)  grant ST/J001651/1
and from the Spanish Ministry  of Economy and Competitiveness (MINECO)
grant  AYA2011-24728.  EMC  is supported  by Padua  University through
grants  60A02-5857/13, 60A02-5833/14,  60A02-4434/15, and  CPDA133894.
RGD and RGB acknowledge the support from the project AyA2014-57490-P y
JA-FQM-2828.  Support for  LG is provided by the  Ministry of Economy,
Development, and Tourism's Millennium Science Initiative through grant
IC120009 awarded  to The  Millennium Institute of  Astrophysics (MAS),
and CONICYT through FONDECYT  grant 3140566.  RAM acknowledges support
by the Swiss National  Science Foundation.  IIM acknowledges financial
support from the  Spanish AEI and European FEDER  fundings through the
research  project AYA2016-76682-C3-1P.   This paper  is based  on data
from  the  Calar  Alto  Legacy  Integral  Field  Area  Survey,  CALIFA
(http://califa.caha.es), funded  by the  Spanish Ministery  of Science
under    grant   ICTS-2009-10,    and    the   Centro    Astron\'omico
Hispano-Alem\'an.

Based on  observations collected  at the Centro  Astron\'omico Hispano
Alem\'an  (CAHA) at  Calar Alto,  operated jointly  by the  Max-Planck
Institut  f\"ur  Astronomie  and  the Instituto  de  Astrof\'isica  de
Andaluc\'ia (CSIC)




\bibliographystyle{mnras}
\bibliography{reference} 


\onecolumn
\appendix
\section{Structural parameters of the galaxy sample}

\begin{center}
\begin{longtable}{p{2.3cm} p{1.4cm} p{1.4cm} p{1.4cm} p{1.4cm} p{1.5cm} p{2cm} p{1.5cm}}
\caption{\label{tab:decompbulge}Structural parameters of the bulge structures in our sample galaxies.}\\     
\hline\hline
\multicolumn{1}{c}{Galaxy} & 
\multicolumn{1}{c}{Band} & 
\multicolumn{1}{c}{$\mu_e$}& 
\multicolumn{1}{c}{$r_{\rm e}$}& 
\multicolumn{1}{c}{$r_{\rm e}$}& 
\multicolumn{1}{c}{$n$}&
\multicolumn{1}{c}{$q_{\rm bulge}$}& 
\multicolumn{1}{c}{$PA_{\rm bulge}$}\\ 
\multicolumn{1}{c}{}&
\multicolumn{1}{c}{}& 
\multicolumn{1}{c}{(mag/\arcsec$^2$)}& 
\multicolumn{1}{c}{(\arcsec)}&
\multicolumn{1}{c}{(kpc)}& 
\multicolumn{1}{c}{}&
\multicolumn{1}{c}{}& 
\multicolumn{1}{c}{($^{\circ}$)}\\ 
\multicolumn{1}{c}{(1)}& 
\multicolumn{1}{c}{(2)}& 
\multicolumn{1}{c}{(3)}&
\multicolumn{1}{c}{(4)}& 
\multicolumn{1}{c}{(5)}&
\multicolumn{1}{c}{(6)}& 
\multicolumn{1}{c}{(7)}&
\multicolumn{1}{c}{(8)}\\ 
\hline
\endfirsthead
\caption{continued.}\\
\hline\hline
\multicolumn{1}{c}{Galaxy} &
\multicolumn{1}{c}{Band} & 
\multicolumn{1}{c}{$\mu_e$}& 
\multicolumn{1}{c}{$r_{\rm e}$}& 
\multicolumn{1}{c}{$r_{\rm e}$}&
\multicolumn{1}{c}{$n$}&
\multicolumn{1}{c}{$q_{\rm bulge}$}& 
\multicolumn{1}{c}{$PA_{\rm bulge}$}\\ 
\multicolumn{1}{c}{}& 
\multicolumn{1}{c}{}&
\multicolumn{1}{c}{(mag/\arcsec$^2$)}& 
\multicolumn{1}{c}{(\arcsec)}& 
\multicolumn{1}{c}{(kpc)}& 
\multicolumn{1}{c}{}& 
\multicolumn{1}{c}{}& 
\multicolumn{1}{c}{($^{\circ}$)}\\ 
\multicolumn{1}{c}{(1)}& 
\multicolumn{1}{c}{(2)}& 
\multicolumn{1}{c}{(3)}& 
\multicolumn{1}{c}{(4)}& 
\multicolumn{1}{c}{(5)}& 
\multicolumn{1}{c}{(6)}&
\multicolumn{1}{c}{(7)}&
\multicolumn{1}{c}{(8)}\\
\hline
\endhead
\hline
\endfoot
IC2341           &   $g$  &  19.6$\pm$0.2 &  2.9$\pm$0.3 &  1.0$\pm$0.1  &   1.3$\pm$0.1  &  0.52$\pm$0.03  &     7$\pm$4   \\
IC2341           &   $r$  &  18.9$\pm$0.2 &  3.0$\pm$0.3 &  1.0$\pm$0.1  &   1.4$\pm$0.1  &  0.55$\pm$0.03  &     5$\pm$4   \\
IC2341           &   $i$  &  18.5$\pm$0.2 &  3.0$\pm$0.3 &  1.0$\pm$0.1  &   1.5$\pm$0.1  &  0.54$\pm$0.03  &     5$\pm$4   \\
MCG-01-52-012    &   $g$  &  19.2$\pm$0.2 &  3.5$\pm$0.4 &  0.9$\pm$0.1  &   1.6$\pm$0.1  &  0.48$\pm$0.01  &     86$\pm$3   \\
MCG-01-52-012    &   $r$  &  19.4$\pm$0.2 &  3.3$\pm$0.4 &  0.8$\pm$0.1  &   1.6$\pm$0.1  &  0.46$\pm$0.01  &     85$\pm$3   \\
MCG-01-52-012    &   $i$  &  18.6$\pm$0.2 &  3.5$\pm$0.4 &  0.9$\pm$0.1  &   1.7$\pm$0.1  &  0.46$\pm$0.01  &     86$\pm$3   \\
NGC0364          &   $g$  &  19.0$\pm$0.3 &  1.5$\pm$0.3 &  0.5$\pm$0.1  &   1.4$\pm$0.2  &  0.85$\pm$0.05  &     42$\pm$6   \\
NGC0364          &   $r$  &  18.3$\pm$0.3 &  1.7$\pm$0.3 &  0.5$\pm$0.1  &   1.4$\pm$0.2  &  0.90$\pm$0.05  &     41$\pm$6   \\
\end{longtable}
\end{center}

\clearpage
\begin{small}
\begin{center}
\begin{longtable}{p{2.1cm} p{0.3cm} p{1.3cm} p{1.3cm} p{1.3cm} p{1.1cm} p{1.3cm} p{1.3cm} p{1.2cm} p{1.5cm} p{1.2cm}}
\caption{\label{tab:decompdisc}Structural parameters of the disc structures in our sample galaxies.}\\     
\hline\hline
\multicolumn{1}{c}{Galaxy} & 
\multicolumn{1}{c}{Band} &
\multicolumn{1}{c}{$\mu_0$}& 
\multicolumn{1}{c}{$h$}& 
\multicolumn{1}{c}{$h$}& 
\multicolumn{1}{c}{$r_{break}$}& 
\multicolumn{1}{c}{$r_{break}$}&
\multicolumn{1}{c}{$h_0$}& 
\multicolumn{1}{c}{$h_0$}&
\multicolumn{1}{c}{$q_{\rm disc}$}& 
\multicolumn{1}{c}{$PA_{\rm disc}$}\\ 
\multicolumn{1}{c}{}& 
\multicolumn{1}{c}{}&
\multicolumn{1}{c}{(mag/\arcsec$^2$)}& 
\multicolumn{1}{c}{(\arcsec)}& 
\multicolumn{1}{c}{(kpc)}& 
\multicolumn{1}{c}{(\arcsec)}& 
\multicolumn{1}{c}{(kpc)}& 
\multicolumn{1}{c}{(\arcsec)}& 
\multicolumn{1}{c}{(kpc)}&
\multicolumn{1}{c}{}&
\multicolumn{1}{c}{($^{\circ}$)}\\ 
\multicolumn{1}{c}{(1)}& 
\multicolumn{1}{c}{(2)}& 
\multicolumn{1}{c}{(3)}& 
\multicolumn{1}{c}{(4)}&
\multicolumn{1}{c}{(5)}& 
\multicolumn{1}{c}{(6)}& 
\multicolumn{1}{c}{(7)}&
\multicolumn{1}{c}{(8)}&
\multicolumn{1}{c}{(9)}&
\multicolumn{1}{c}{(10)}&
\multicolumn{1}{c}{(11)}\\ 
\hline
\endfirsthead
\caption{continued.}\\
\hline\hline
\multicolumn{1}{c}{Galaxy} & 
\multicolumn{1}{c}{Band} &
\multicolumn{1}{c}{$\mu_0$}& 
\multicolumn{1}{c}{$h$}& 
\multicolumn{1}{c}{$h$}& 
\multicolumn{1}{c}{$r_{break}$}& 
\multicolumn{1}{c}{$r_{break}$}& 
\multicolumn{1}{c}{$h_0$}& 
\multicolumn{1}{c}{$h_0$}&
\multicolumn{1}{c}{$q_{\rm disc}$}& 
\multicolumn{1}{c}{$PA_{\rm disc}$}\\  
\multicolumn{1}{c}{}&
\multicolumn{1}{c}{}&
\multicolumn{1}{c}{(mag/\arcsec$^2$)}& 
\multicolumn{1}{c}{(\arcsec)}& 
\multicolumn{1}{c}{(kpc)}& 
\multicolumn{1}{c}{(\arcsec)}& 
\multicolumn{1}{c}{(kpc)}& 
\multicolumn{1}{c}{(\arcsec)}& 
\multicolumn{1}{c}{(kpc)}&
\multicolumn{1}{c}{}& 
\multicolumn{1}{c}{($^{\circ}$)}\\ 
\multicolumn{1}{c}{(1)}& 
\multicolumn{1}{c}{(2)}& 
\multicolumn{1}{c}{(3)}& 
\multicolumn{1}{c}{(4)}& 
\multicolumn{1}{c}{(5)}& 
\multicolumn{1}{c}{(6)}& 
\multicolumn{1}{c}{(7)}&
\multicolumn{1}{c}{(8)}&
\multicolumn{1}{c}{(9)}&
\multicolumn{1}{c}{(10)}&
\multicolumn{1}{c}{(11)}\\
\hline
\endhead
\hline
\endfoot
MCG-01-52-012    &   $g$  &    20.5$\pm$0.2 &   8.1$\pm$0.5 &    2.0$\pm$0.1  &  -                &      -            &  -          &       -        & 0.79$\pm$0.03  &     44$\pm$4    \\
MCG-01-52-012    &   $r$  &    20.7$\pm$0.2 &   8.0$\pm$0.5 &    2.0$\pm$0.1  &  -                &      -            &  -          &       -        & 0.79$\pm$0.03  &     47$\pm$4    \\
MCG-01-52-012    &   $i$  &    19.8$\pm$0.2 &   7.8$\pm$0.5 &    1.9$\pm$0.1  &  -                &      -            &  -          &       -        & 0.80$\pm$0.03  &     43$\pm$4    \\
IC2341           &   $g$  &    20.6$\pm$0.1 &   9.7$\pm$0.4 &    3.3$\pm$0.1  &   32$\pm$2        &    11.1$\pm$0.6   &   14$\pm$2  &   4.9$\pm$0.6  & 0.52$\pm$0.01  &     2$\pm$2   \\
IC2341           &   $r$  &    19.9$\pm$0.1 &   9.9$\pm$0.4 &    3.4$\pm$0.1  &   33$\pm$2        &    11.2$\pm$0.6   &   15$\pm$2  &   5.4$\pm$0.7  & 0.52$\pm$0.01  &     2$\pm$2    \\
IC2341           &   $i$  &    19.6$\pm$0.1 &   10.0$\pm$0.4 &   3.4$\pm$0.1  &   32$\pm$2        &    11.0$\pm$0.6   &   15$\pm$2  &   5.2$\pm$0.7  & 0.53$\pm$0.01  &     2$\pm$2    \\
NGC0364          &   $g$  &    21.1$\pm$0.2 &   12.6$\pm$0.5 &   4.0$\pm$0.2  &  -                &  -                &  -          &       -        & 0.72$\pm$0.02  &     33$\pm$2    \\
NGC0364          &   $r$  &    20.3$\pm$0.2 &   12.6$\pm$0.5 &   4.0$\pm$0.2  &  -                &  -                &  -          &       -        & 0.72$\pm$0.02  &     33$\pm$2    \\
\end{longtable}
\end{center}
\end{small}
\clearpage
\begin{center}
\begin{longtable}{p{2.4cm} p{1.4cm} p{1.4cm} p{1.4cm} p{1.4cm} p{2cm} p{1.5cm}}
\caption{\label{tab:decompbar}Structural parameters of the bar structures in our sample galaxies.}\\     
\hline\hline
\multicolumn{1}{c}{Galaxy} &  
\multicolumn{1}{c}{Band} &
\multicolumn{1}{c}{$\mu_{bar}$}& 
\multicolumn{1}{c}{$a_{\rm bar}$}& 
\multicolumn{1}{c}{$a_{\rm bar}$}& 
\multicolumn{1}{c}{$q_{\rm bar}$}&
\multicolumn{1}{c}{$PA_{\rm bar}$}\\ 
\multicolumn{1}{c}{}&
\multicolumn{1}{c}{}&
\multicolumn{1}{c}{(mag/\arcsec$^2$)}& 
\multicolumn{1}{c}{(\arcsec)}& 
\multicolumn{1}{c}{(kpc)}& 
\multicolumn{1}{c}{}& 
\multicolumn{1}{c}{($^{\circ}$)}\\ 
\multicolumn{1}{c}{(1)}& 
\multicolumn{1}{c}{(2)}& 
\multicolumn{1}{c}{(3)}& 
\multicolumn{1}{c}{(4)}&
\multicolumn{1}{c}{(5)}&
\multicolumn{1}{c}{(6)}&
\multicolumn{1}{c}{(7)}\\ 
\hline
\endfirsthead
\caption{continued.}\\
\hline\hline
\multicolumn{1}{c}{Galaxy} & 
\multicolumn{1}{c}{Band} & 
\multicolumn{1}{c}{$\mu_{bar}$}& 
\multicolumn{1}{c}{$a_{\rm bar}$}& 
\multicolumn{1}{c}{$a_{\rm bar}$}& 
\multicolumn{1}{c}{$q_{\rm bar}$}&
\multicolumn{1}{c}{$PA_{\rm bar}$}\\ 
\multicolumn{1}{c}{}&
\multicolumn{1}{c}{}&
\multicolumn{1}{c}{(mag/\arcsec$^2$)}& 
\multicolumn{1}{c}{(\arcsec)}& 
\multicolumn{1}{c}{(kpc)}& 
\multicolumn{1}{c}{}& 
\multicolumn{1}{c}{($^{\circ}$)}\\ 
\multicolumn{1}{c}{(1)}& 
\multicolumn{1}{c}{(2)}& 
\multicolumn{1}{c}{(3)}& 
\multicolumn{1}{c}{(4)}& 
\multicolumn{1}{c}{(5)}&
\multicolumn{1}{c}{(6)}&
\multicolumn{1}{c}{(7)}\\
\hline
\endhead
\hline
\endfoot
IC2341           &   $g$  &    -              &   -              &      -          &  -              &    -  \\
IC2341           &   $r$  &    -              &   -              &      -          &  -              &    -  \\
IC2341           &   $i$  &    -              &   -              &      -          &  -              &    -  \\
MCG-01-52-012    &   $g$  &    -              &   -              &     -           &  -              &    -  \\
MCG-01-52-012    &   $r$  &    -              &   -              &     -           &  -              &    -  \\
MCG-01-52-012    &   $i$  &    -              &   -              &     -           &  -              &    -  \\
NGC0364          &   $g$  &     20.9$\pm$0.4  &    12.6$\pm$1.2  &    4.0$\pm$0.1  &  0.55$\pm$0.06  &     91$\pm$4  \\
NGC0364          &   $r$  &     20.0$\pm$0.4  &    12.7$\pm$1.2  &    4.0$\pm$0.1  &  0.55$\pm$0.06  &     91$\pm$4  \\
\end{longtable}
\end{center}

\twocolumn
\section{Effects of pixelization and PSF on the stellar kinematics}
\label{sec:psf}

Despite  the  good  spatial  coverage   and  sampling  of  the  CALIFA
datacubes, we face the problem that  some of our bulges have effective
radii comparable  to the  measured PSF of  the final  CALIFA datacubes
\citep[$\sim$2.5 arcsec;][]{garciabenito15}.

   \begin{figure*}
     \begin{center}
   \includegraphics[]{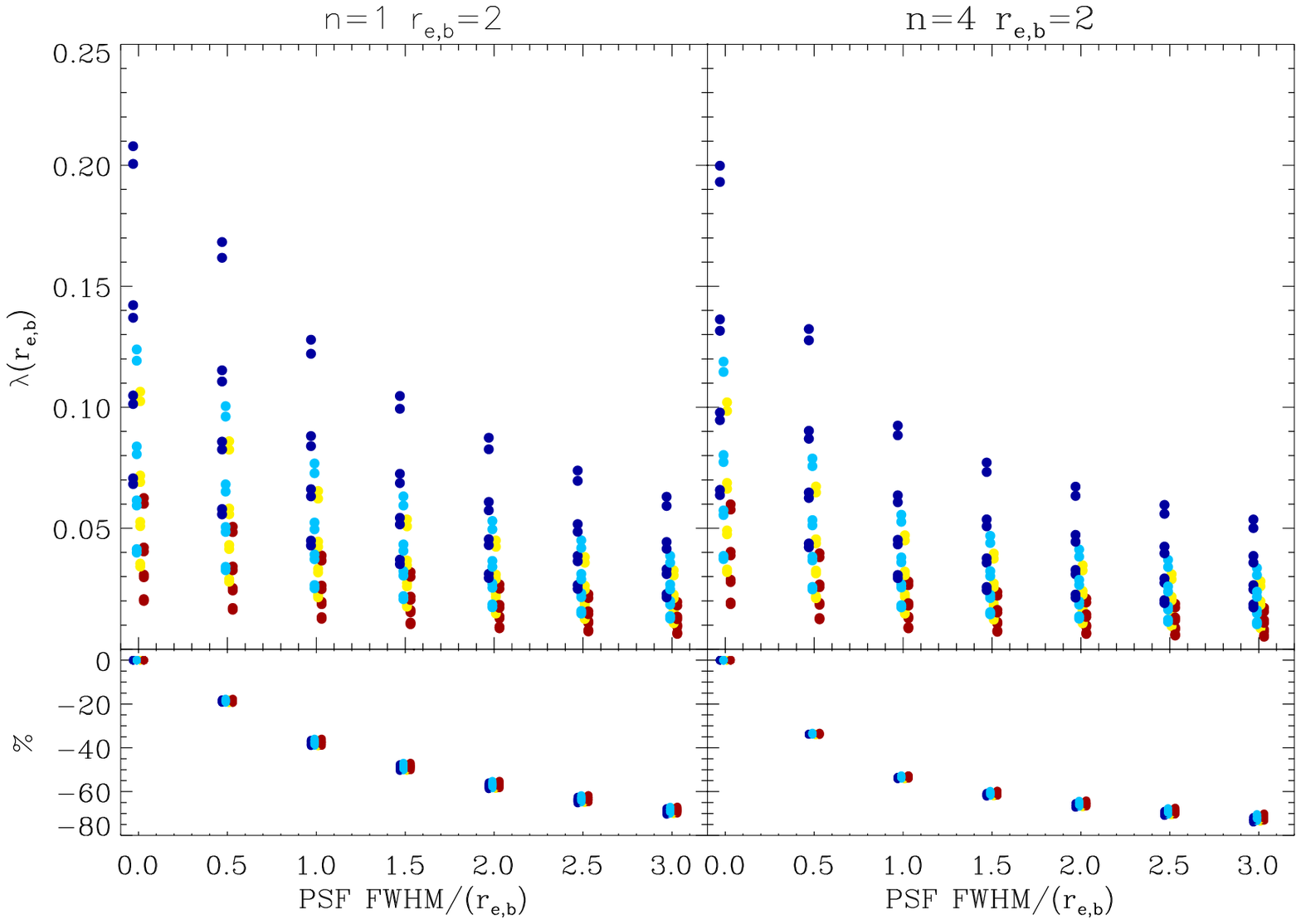}
      \caption{{\it   Upper   panels}.    Distribution   of   measured
        $\lambda_{e,  b}$ as  a function  of the  PSF/$r_{e,b}$ ratio.
        Different colours represent  different combinations of $v_{\rm
          max}$ and $\sigma_{\rm max}$.  Points have been artificially
        shifted in the x-axis for representation purposes.  {\it Lower
          panels}.   Distribution  of $\lambda_{e,  b}$
        differences  in percentages  with respect  to the  models with
        PSF/$r_{e,b}$ =0, i.e., the ideal  case. Left and right panels
        show  all possible  combinations of  the kinematic  parameters
        with $r_{e,b}$=2 and $n=1$ and $n=4$, respectively.}
        \label{fig:fixsbd}
     \end{center}
   \end{figure*}


In  order  to  estimate  the  uncertainties  due  to  the  effects  of
pixelization and PSF, and eventually  include them in the error budget
of our  integrated kinematic measurements,  we decided to carry  out a
set of tests using simulations with mock datacubes.  To this aim, mock
datacubes  were created  from scratch  including both  photometric and
kinematics properties comparable to those of our sample galaxies. Mock
datacubes also share  the technical properties of real  CALIFA data in
terms of spatial and spectral resolution.

In  detail,  the  mock  datacubes  were  created  with  the  following
properties:  i)   we  used  a  single   stellar  population  spectrum,
corresponding to  an old and  metal rich  star, chosen from  the MILES
library \citep{sanchezblazquez06}  to create  our datacube.   Since we
are only  interested in the  effect of  the spatial resolution  on the
kinematic  measurements, a  realistic  combination  of single  stellar
populations (SSPs) to  reproduce a galaxy is  not necessary. Therefore
all the  spaxels in the  datacube were assumed  to have the  same SSP.
ii)  We used  a set  of SBD  to reproduce  realistic spaxel  intensity
variation within  the datacube.  Each  spaxel intensity was  scaled to
follows  a SBD  described by  a S\'ersic  profile with  values of  the
effective radius $r_{e}=1,  2,$ and 5 arcsec and  S\'ersic index $n=1,
2, 3,$ and  4.  Then, all datacubes were convolved  with values of the
FWHM$=0, 1,  2, 3, 4,  5,$ and 6 arcsecs  where FWHM =0  represent the
perfect model.  iii) The velocity field was modelled assuming that the
mock galaxies have  a rotation curve that  follow the parameterisation
by \citet{salucci07}

\begin{equation} 
v_c (r) = v_{\rm max}\frac{r}{\sqrt{r^2_{v} + r^2}} 
\label{eqn:rotvel} 
\end{equation} 
%
where  $v_{\rm max}$  is  the maximum  rotation  velocity and  $r_{v}$
define the spatial  rising of the rotation velocity  profile.   We
  projected the  velocity field  on the  sky plane  assuming cartesian
  coordinates with  the origin in  the centre of the  galaxy, $x-$axis
  aligned along  the apparent major  axis of the galaxy,  and $z-$axis
  along the line of sight directed towards the observer. The sky plane
  is confined to the ($x, y$) plane.  If the galaxy has an inclination
  angle i (with i = 0$^{\circ}$ corresponding to the face-on case), at
  a given sky  point with coordinates ($x, y$),  the observed velocity
  $v(x, y)$ is

\begin{equation} 
v (x,y) = v_{c}(x,y)\, \sin{i} \, \cos{\phi}
\label{eqn:rotvel2} 
\end{equation} 
where $\phi$ is  the azimuthal angle measured from  the apparent major
axis  of the  galaxy.  The spectra  in each  spaxel  was then  shifted
according to  this rotation curve  using values of $v_{\rm  max}=$ 150
and  300 km/s,  and $r_{v}=$  10 and  15 arcsec.   The velocity  field
parameterisation was  tested against the  real data.  To this  aim, we
derived  the rotation  curves from  our galaxy  sample using  the {\it
  kinemetry}  routine  developed  by \citet{krajnovic06}.   Then,  the
rotation  curves   were  fitted   using  Eqs.    \ref{eqn:rotvel}  and
\ref{eqn:rotvel2} with  $v_{\rm max}$ and $r_{v}$  as free parameters.
The mock  galaxy values correspond  to the minimum and  maximum values
obtained from this fit.  The  velocity dispersion of the real galaxies
was modelled by a simple exponentially declining profile with two free
parameters: the  maximum velocity dispersion ($\sigma_{\rm  max}$) and
the scale-length ($r_{\sigma}$).  Attending  to the typical values for
our  galaxy sample  (obtained  from  the fit  to  the {\it  kinemetry}
velocity dispersion profiles)  we built the datacubes  with a velocity
dispersion  profile  with  $\sigma_{\rm  max}=$150 and  200  km/s  and
$r_{\sigma}=$20 and  40 arcsecs.  Finally, all datacubes  were created
using models with two different  inclinations $i=$30 and 60 degrees to
test possible inclination effects on our measurements.  A final sample
of 2688 mock datacubes was created.

Since measurement  errors were  already included  in the  error budget
using the  observed galaxy  datacubes (see  Sect.  \ref{sec:kinprop}),
mock  datacubes were  created  noise free.   Furthermore, to  minimise
problems related to  template mismatching, we decided to  run the pPXF
algorithm   using  the   same  SSP   as  that   used  to   create  the
datacubes. Then, the pPXF algorithm was  run as for real datacubes and
the stellar  kinematic maps of  velocity and velocity  dispersion were
obtained and analysed as described in Sect. \ref{sec:kinematics}.

We measured  the values of the  $(v/\sigma)_{e,b}$ and $\lambda_{e,b}$
using        Eqs.~\ref{eqn:vsigma}        and        \ref{eqn:lambda}.
Figure~\ref{fig:fixsbd} shows, for  a given photometric configuration,
an example  of the  biases introduced when  the measurement  radii are
comparable to the spatial resolution  of the data.  The measured value
of $\lambda_{e,b}$  is a strong  function of the  PSF/$r_{e,b}$ ratio,
with differences reaching up to 70\% from the actual value, and always
affecting the measurements towards lower values of $\lambda_{e,b}$. In
fact,  pixelation and  PSF effect  can be  considered as  a systematic
variation and therefore  a correction factor should be  applied to the
data.  Fortunately, Figure~\ref{fig:fixsbd} (bottom panels) also shows
that the correction factor does  not strongly depend on the kinematics
of the  system but mostly  on their photometric configuration  and the
PSF/$r_{e,b}$  ratio.   In  fact,  the errors  due  to  the  different
kinematic of the system  (different colours in Figure~\ref{fig:fixsbd})
are  always  much  smaller  ($\le10\%$)  than  the  correction  factor
itself. Therefore,  since the  photometric parameters can  be obtained
with a better spatial resolution (SDSS) than the datacubes (CALIFA) we
can  safely  compute  the  correction factor,  and  its  corresponding
dispersion,   and   properly   correct  the   integrated   values   of
$(v/\sigma)_{e,b}$ and $\lambda_{e,b}$.

   \begin{figure}
     \begin{center}
   \includegraphics[width=0.43\textwidth]{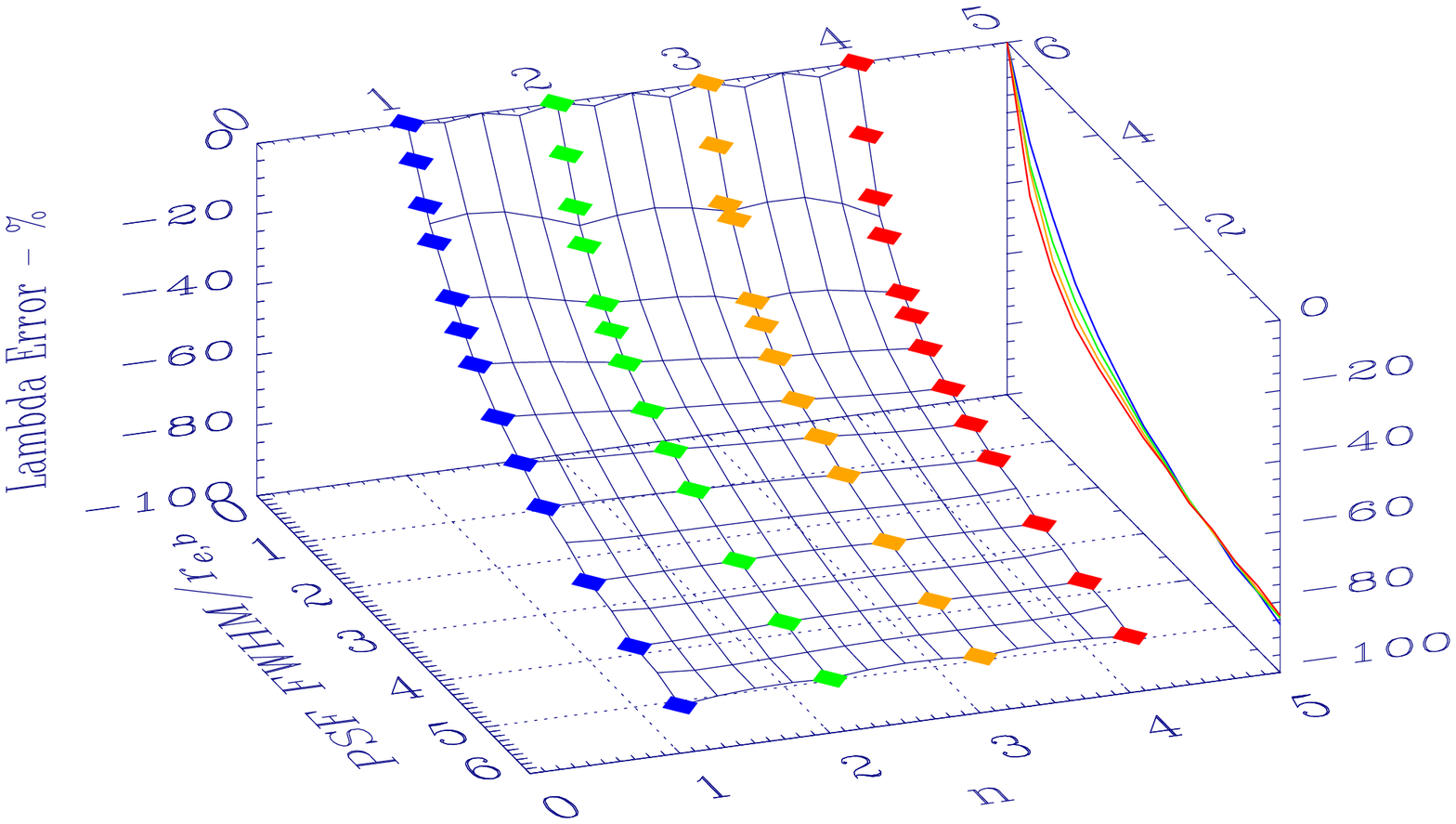}
      \caption{Distribution of  the $\lambda_{e,b}$  correction factor
        in   the  PSF/$r_{e,b}$   ratio  and   S\'ersic  index   ($n$)
        plane. Each point represents the median value of the different
        models.   Different colours  represents  models with  different
        values of  the S\'ersic index  $n$.  The surface  resulting of
        interpolating the correction factor  is also shown. The result
        of this  interpolation for  values of  $n$=1, 2,  3, and  4 is
        shown in the y-z plane.}
        \label{fig:interp}
     \end{center}
   \end{figure}


We computed the correction factor for each galaxy by interpolating the
plane shown in Figure~\ref{fig:interp}  using the corresponding values
of $r_{e,b}$/PSF  ratio and  S\'ersic index ($n$)  for each  bulge. In
order  to  account  for  the  errors due  to  the  different  possible
kinematics, inclinations,  and bulge  effective radius  ($r_e$) (shown
with different  colours in  Figure~\ref{fig:fixsbd}), we  perform this
interpolation in a Monte Carlo  fashion. We created 2000 interpolation
planes by  randomly changing the position  of every {\it node}  of the
plane, i.e., for all the  different kinematic configurations. Both the
mean and  standard deviation were  taken as the correction  factor and
its  error, respectively.   This error  was then  propagated into  the
value of  $\lambda_{e,b}$ and  summed quadratically  with measurements
errors as described in Sect.~\ref{sec:rotsupport}.


\bsp	
\label{lastpage}
\end{document}